\documentclass{article}
\usepackage[utf8]{inputenc}
\usepackage{booktabs}
\usepackage{a4wide}
\usepackage{multirow}
\usepackage[table]{xcolor}
\usepackage[hyphens]{url}
\usepackage{palatino}
\usepackage{graphicx}
\usepackage[utf8]{inputenc}
\usepackage{authblk}

\begin{document}
\title{Analysing the fall 2020 Emotet campaign}
\date{}
\author[1]{Constantinos Patsakis}
\author[2]{Anargyros Chrysanthou}

\affil[1]{University of Piraeus and Athena Research Center}
\affil[2]{Neurosoft}
\maketitle

\section{Introduction}
\addcontentsline{toc}{section}{Introduction}
Emotet is a modular trojan that is known since 2014 \cite{salvio2014new} and is launching various campaigns from time to time. Like in previously observed campaigns, after infecting its victims, Emotet further infects its victims with other malware (information stealers, ransomware, etc.). In general, Emotet has tight bonds with several other known malware, such as Ursnif, Dridex and BitPayme as they share the same loader \cite{tm2018}, as well as Qbot \cite{qbot} and Trickbot\cite{cybereason}. In fact, in the past few years, Emotet acts more as a malware loader with various capabilities that distinguish it from its peers \cite{wifi_module}. During the latest campaign, Emotet is further infecting its victims with TrickBot and Ryuk \cite{emo_trick}. The above indicates that Emotet is monetising following the \textit{Malware-as-a-Service (MaaS)} or \textit{Access-as-a-Service (AaaS)} model, namely that Emotet controllers lease Emotet infected devices to malicious parties, in order for the latter to carry out further cyber attacks.

Currently, Emotet is reported by both Member States and private sector respondents as the top malware threat, which affects the EU  \cite{europol2020internet}. The latest campaign of Emotet is based on the well-known \textit{e-mail thread hijacking} method. In essence, the adversary uses legitimate previous e-mail messages, stolen from compromised email clients, to spread to the intended victims. The adversary masquerades malicious e-mails as originating from a legitimate user and sends them to recent e-mail recipients, which are connected to the user. In this way, the adversary significantly increases the chances of one of the potential victims opening the attached document and subsequently getting infected. In the campaign, the attached document is a malicious Microsoft Word file (*.doc), which in the campaign's latest activity is delivered through an encrypted compressed (.zip) file, whose password is depicted in the email body. Actually, the Microsoft Word document was delivered, in the analysed campaign, in three ways, namely (a) a document directly attached in the e-mail, (b) a URL link contained in the e-mail body and (c) within an encrypted compressed file attached in the e-mail.

\begin{figure}[!ht]
    \centering
    \includegraphics[width=\textwidth]{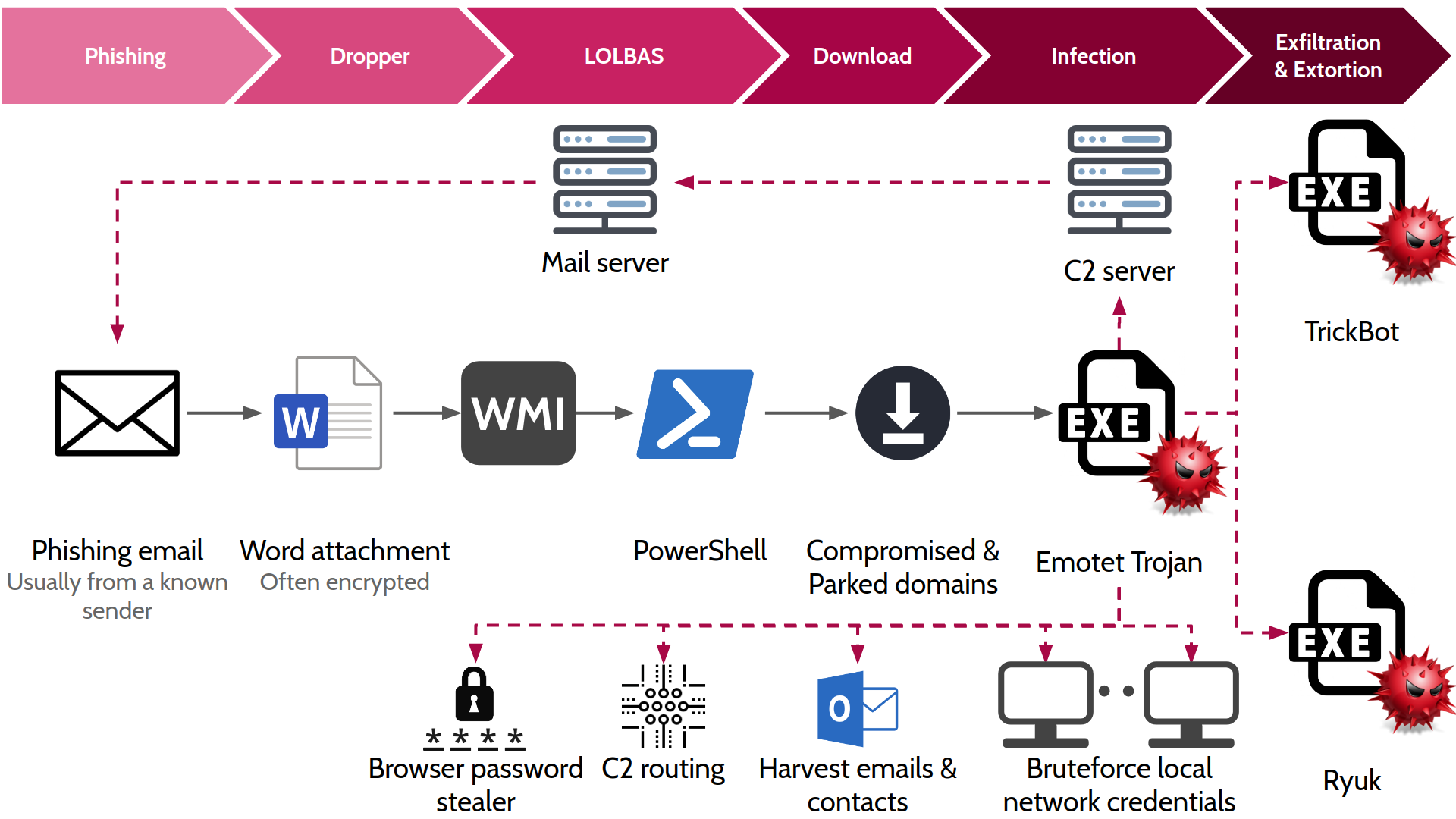}
    \caption{Emotet's latest campaign modus operandi}
    \label{fig:modus}
\end{figure}

The weaponised Microsoft Word file downloads malicious executables from various URLs. These executables are used to attack computers in the same local network, exfiltrate data from the compromised host, and download other ``affiliated'' malware, such as Trickbot and Ryuk, to further spread the infection in the host. Moreover, further e-mail messages are collected and sent to the C2 server, which will be utilised later on in attempts to infect more unsuspected users. Figure \ref{fig:modus} depicts Emotet's modus operandi.

Emotet has a set of individual modules for escalating privileges, harvesting contacts and recent e-mails from Outlook, brute-forcing local network credentials and proxying C2 traffic from other infected devices \cite{vbemotet}. Based on these modules, Emotet manages to infect other nodes once it sets foot on a network, it spreads very quickly, and since it cooperates with other malware, the infected devices are exposed to multiple threats.

At the time of writing Emotet is operating on three different \textit{infrastructures}, meaning that they have different C2 servers, distribution methods, payloads, RSA keys and thus are named as \textit{Epoch 1}, \textit{Epoch 2}, and \textit{Epoch 3}. Nevertheless, from time to time, some C2 servers may be reused by different epochs. 

In the past few weeks, there has been a huge spike in Emotet samples \cite{bro}, with the associated spam campaigns targeting users mainly from Greece, Japan and Lithuania \cite{eset}. To this end, in this report, we analyse the latest campaign of Emotet that had a significant impact in several countries worldwide. We leverage the data of a specifically crafted dataset, which contains emails, documents, executables and domains from the latest campaign. The goal is to analyse the attack vector, map the infrastructure used in various stages of the campaign and perform a surface analysis of Emotet's malicious payloads to assess their potential impact.  
Therefore, in the following paragraphs, we discuss the most critical findings from each part of the extracted information.

\section{Dataset}
Our dataset consists of 3048 e-mail headers, 1968 documents, 749 executables and 1375 domains that have been extracted from these malicious documents. The samples have been collected from public feeds, and clients monitored by Neurosoft, while the e-mail headers are from Neurosoft clients and a big Greek financial institution. For the anonymity of the clients, all recipient related information had been removed prior to the analysis of the headers. The bulk of the above information belongs to the past few months, during which Emotet has re-emerged after several months of inactivity.

\section{URL statistics}
From the 1375 URLs that have been used by Emotet malicious documents to download executables, 619 belong to WordPress installations (45\%). In 108 of them, the CGI path has been used. The URLs appear to originate from compromised domains, and in the vast majority, the address is in depth two of the compromised domain. For instance, they have the following form:

\noindent \texttt{http(s)://compromised\_domain.tld/wp-admin/{\color{red}XYZ/malicious\_path}}

\begin{figure}[!th]
    \centering
    \includegraphics[width=\textwidth]{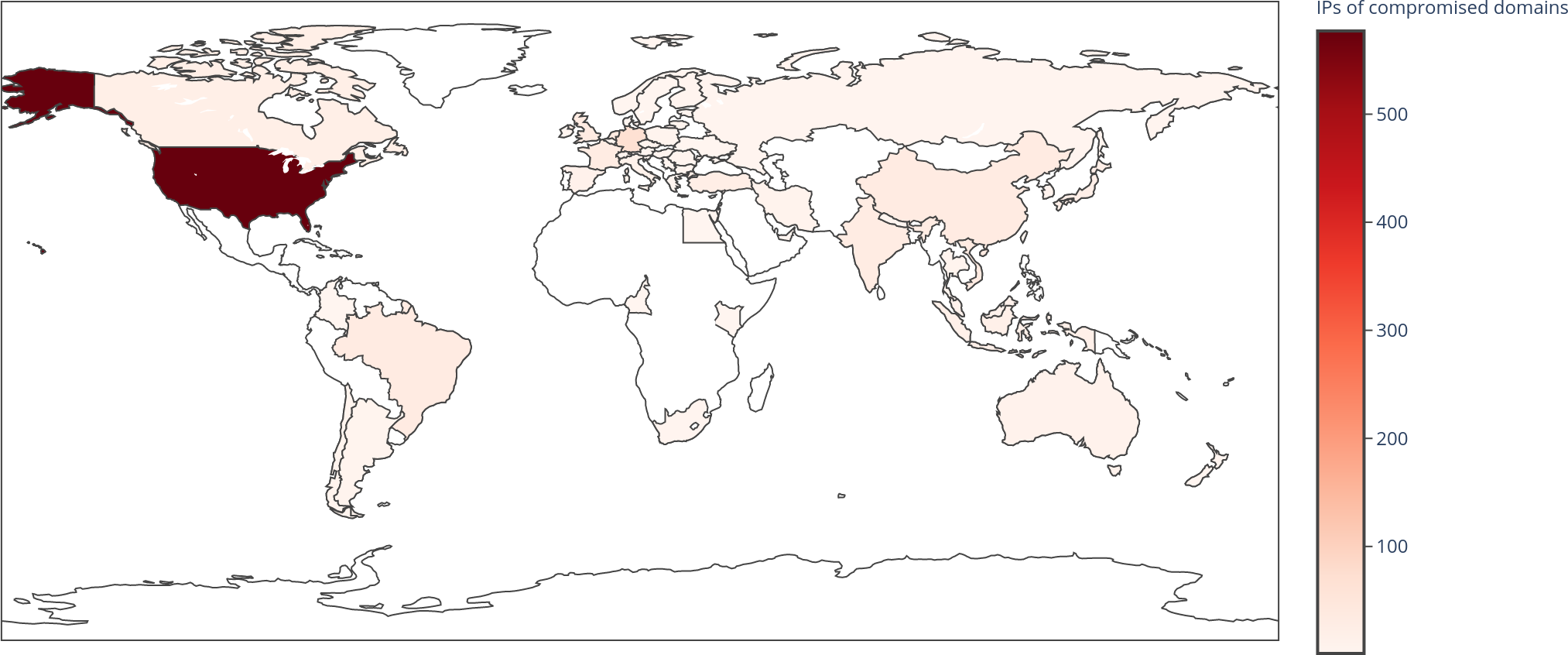}
    \caption{Geolocation of compromised domains.}
    \label{fig:geo_compr}
\end{figure}

These URLs belong to 1276 unique domains, with some of them belonging to the same domain. 1195 of these URLs are still live at the time of writing. These domains correspond to 1129 IP addresses. Figure \ref{fig:geo_compr} illustrates a world map where the IPs of the compromised domains are currently hosted. 

\section{Email statistics}
In our e-mail analysis, we wanted to first determine the actual senders of the emails, as the sender may have used a known to the victim e-mail; however, the reported e-mail is spoofed. Therefore, it is worthwhile to see who is distributing the e-mails, and from which domain. Furthermore, the first hop of the e-mail, which may reveal the actual IP of the sender, is also significant, so the corresponding information was also extracted. Finally, we extracted information relevant to the use of SPF, DKIM, and DMARC.

\begin{figure}
    \centering
    \includegraphics[width=\textwidth]{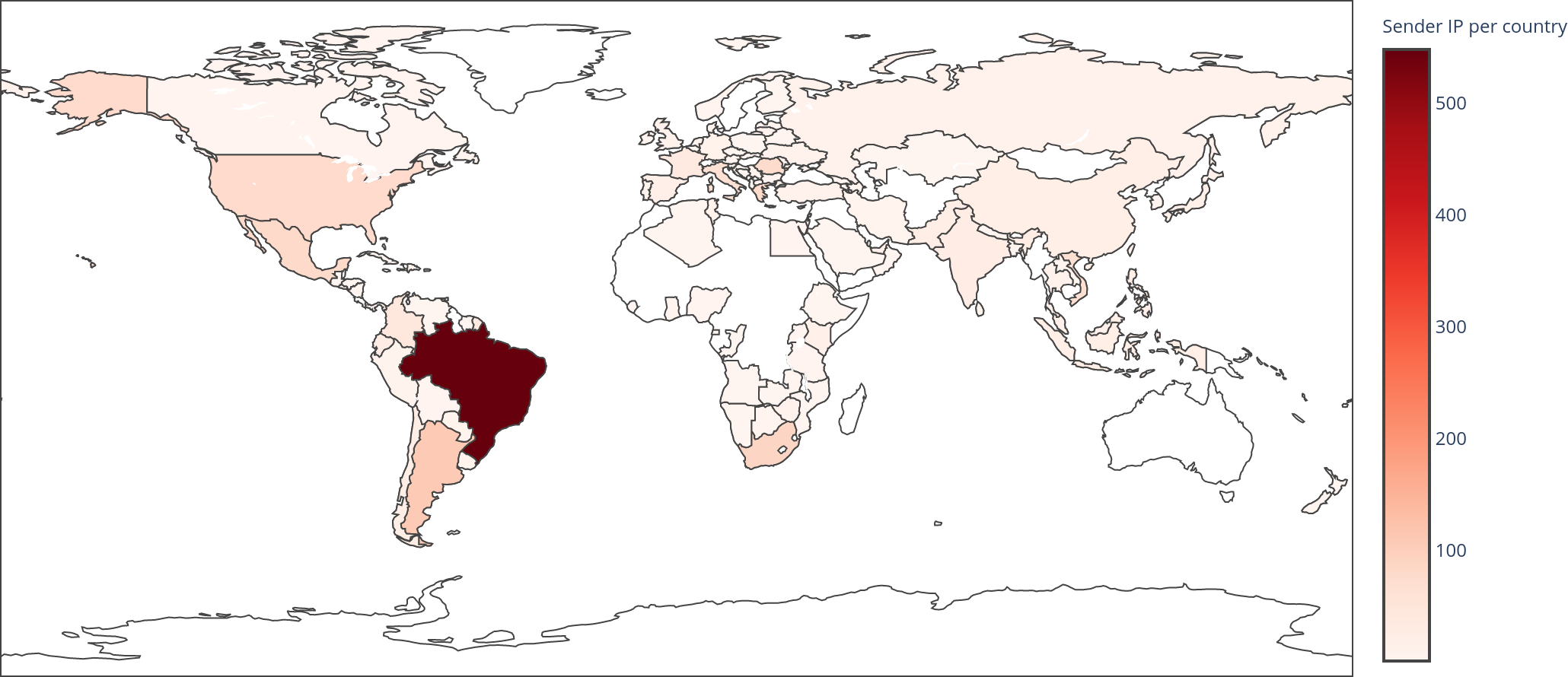}
    \caption{Geolocation of e-mail senders}
    \label{fig:senders}
\end{figure}

The origin of the 1855 e-mail senders that were used to send the spear-phishing e-mails is illustrated in Figure \ref{fig:senders}. Clearly, the biggest part of e-mail senders originates from Brazil, with Latin American countries having by far the biggest share in the distribution of the e-mails.

Digging a bit more into the extracted information, see Table \ref{tab:mail_stats}, we observe that there is no bulk e-mail submission from individual IP addresses. While the emails may have used IP addresses from specific countries, the detected, through the conducted analysis, timezone is mostly -3, which coincides with the timezone of the detected, based on senders' IP address, most used originating country. The latter implies some further spoofing in the emails to hide the true origin of the e-mail senders, which seems to be Brazil.

\begin{table}[!ht]
\centering
\begin{tabular}{lr|lr|lr}
\hline
\textbf{IP} & \textbf{Mails} & \textbf{Timezone} & \textbf{E-mails} & \textbf{IP origin} & \textbf{E-mails} \\ \hline
187.84.237.61 & 24 & -3.0 & 1134 & Brazil & 548 \\
131.196.0.190 & 24 & +2.0 & 532 & Argentina & 110 \\
195.138.93.251 & 21 & +1.0 & 289 & South Africa & 89 \\
78.56.168.37 & 16 & -5.0 & 170 & Mexico & 80 \\
94.103.141.81 & 15 & -6.0 & 154 & Greece & 79 \\
186.4.197.91 & 15 & +7.0 & 143 & United States & 78 \\
128.199.125.199 & 15 & +8.0 & 101 & Romania & 76 \\
84.205.235.9 & 14 & -4.0 & 100 & Vietnam & 70 \\
179.108.2.167 & 14 & +3.0 & 85 & Italy & 66 \\
95.9.220.106 & 13 & +5.5 & 70 & Colombia & 42 \\ \hline
\end{tabular}
\caption{IP of e-mail senders (top 10), detected timezone (top 10), and country of origin of IP address used to send the e-mails (top 10).}
\label{tab:mail_stats}
\end{table}

Some worthwhile statistics about the e-mails is that for instance, in the vast majority of the e-mails there was no SPF check (86.52\%). In the cases where SPF check was performed, only 9.45\% passed it, with the rest of them reporting some errors. The same statistics are also observed in DKIM and DMARK. In DKIM, 86.35\% of the emails did not perform any such check, and only 2.49\% of them passed it, with the rest reporting some error. Finally, in DMARK, 87.14\% did not perform any such check and 8.2\% \textit{passing} (2.36\% pass and 5.84\% bestguesspass -- so no DMARC TXT record for the latter domains exists). 

\section{Dropper statistics}
Emotet dropper comes in the form of a Microsoft Word file that triggers the execution of a VBA macro once the document opens. As in most such scenarios, the VBA code is highly obfuscated, resulting in a call to a LOLBAS \cite{lolbas} that downloads some executable binaries which are subsequently executed. Likewise, Emotet uses PowerShell for its LOLBAS and downloads the binaries from several compromised domains. More precisely, Powershell is actually opened through Windows Management Instrumentation (WMI). As expected, all URLs are obfuscated in both VBA and the PowerShell Payload. In detail, the code contains 7 URLs to be used. Each URL is contacted in sequence. If one of the URLs is live and an executable is downloaded and executed, the subsequent URLs remain unused.

\begin{figure}[!ht]
    \centering
\includegraphics[width=0.45\textwidth]{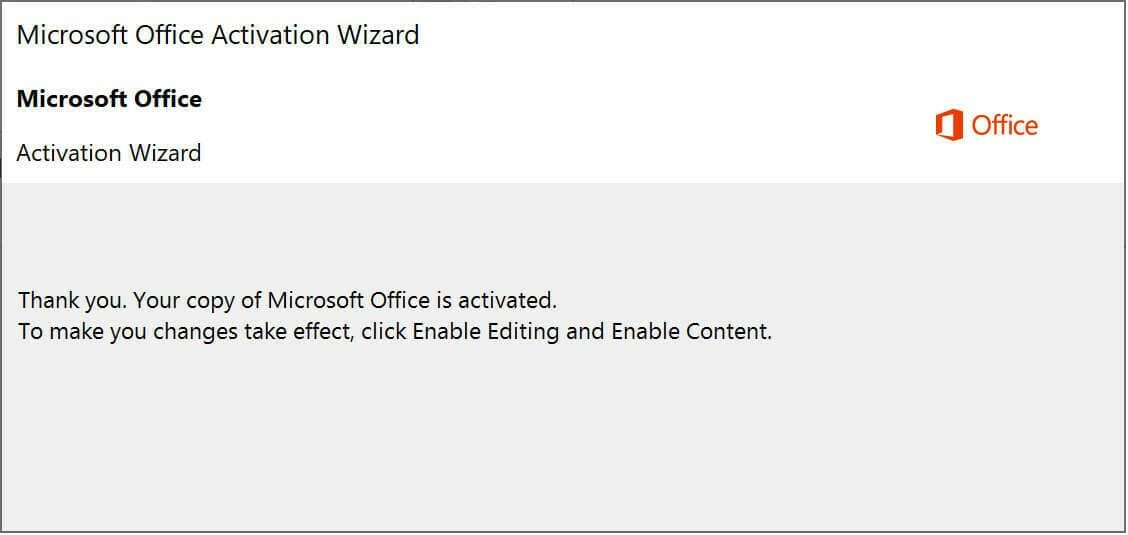}
\includegraphics[width=0.45\textwidth]{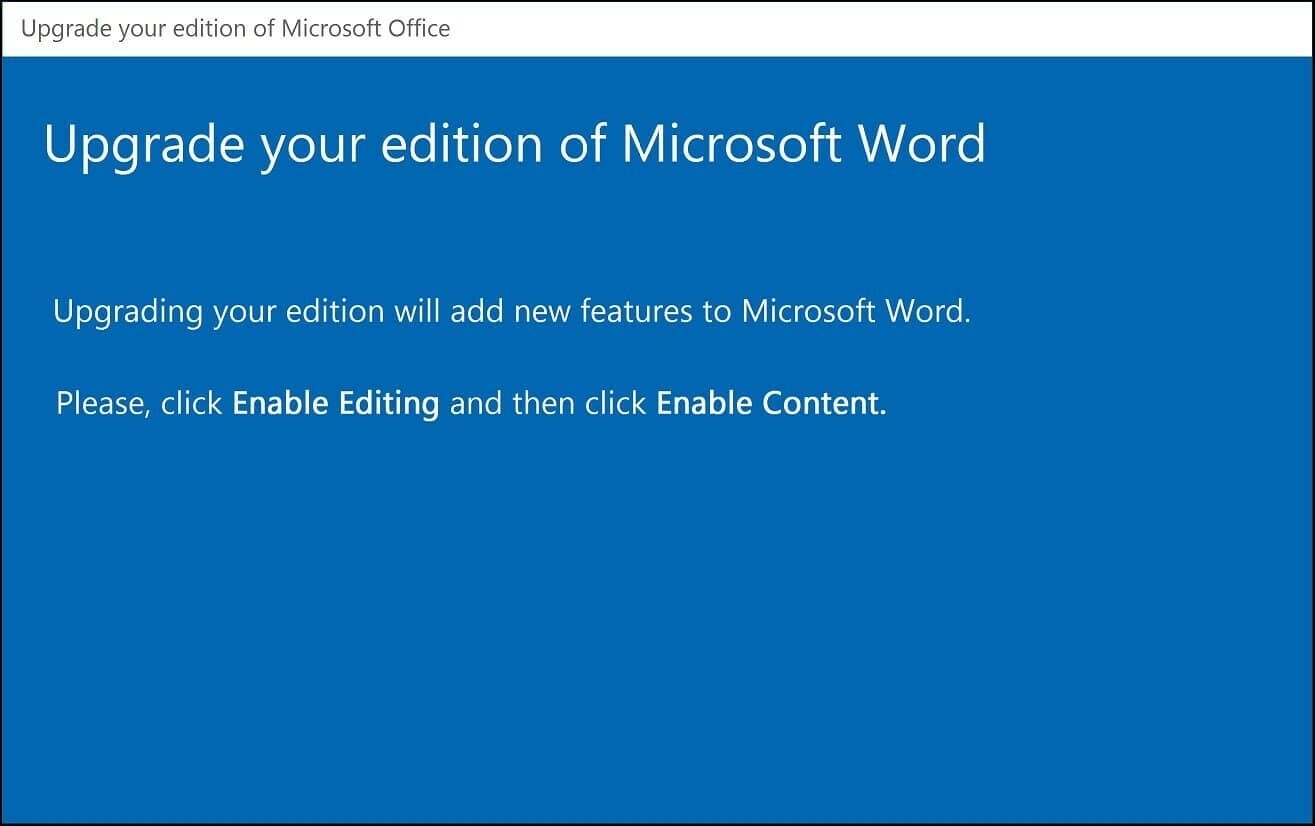}
\includegraphics[width=0.45\textwidth]{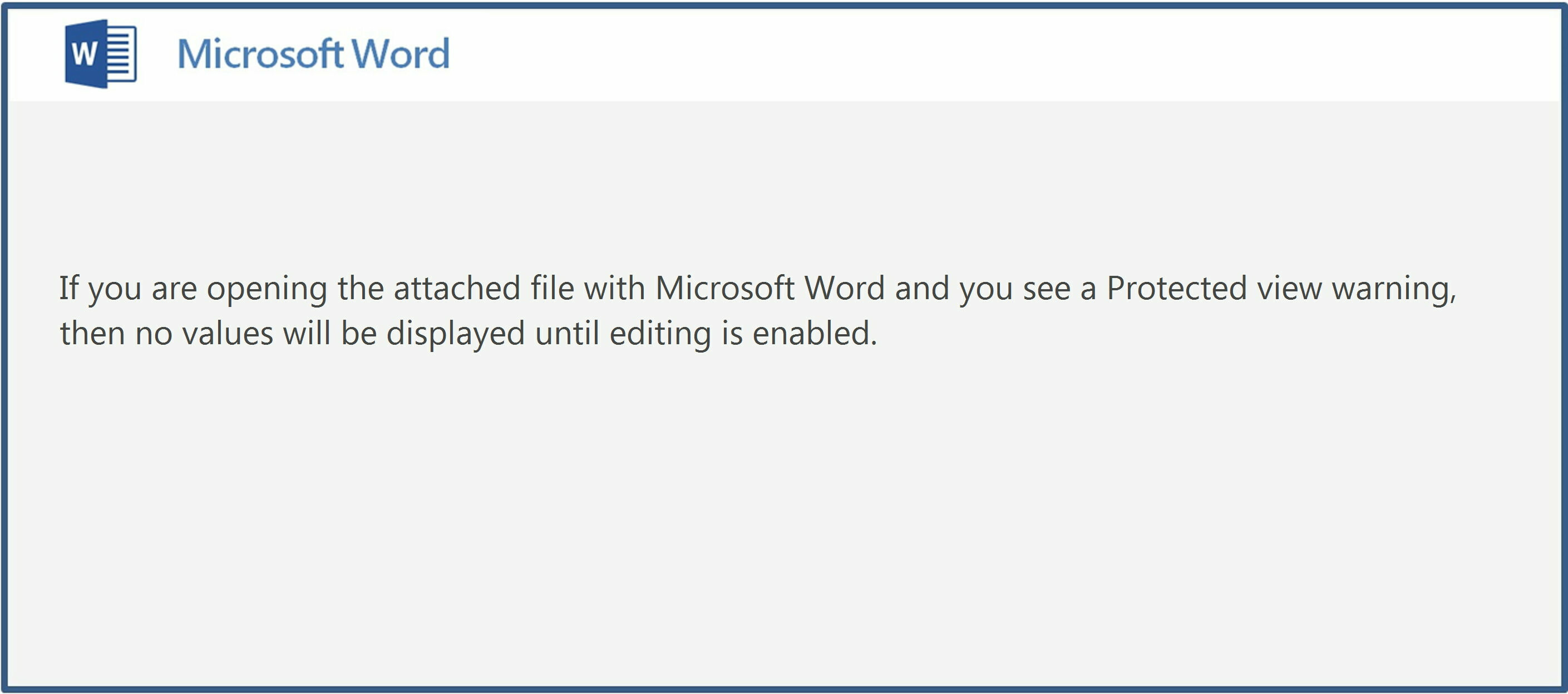}
\includegraphics[width=0.45\textwidth]{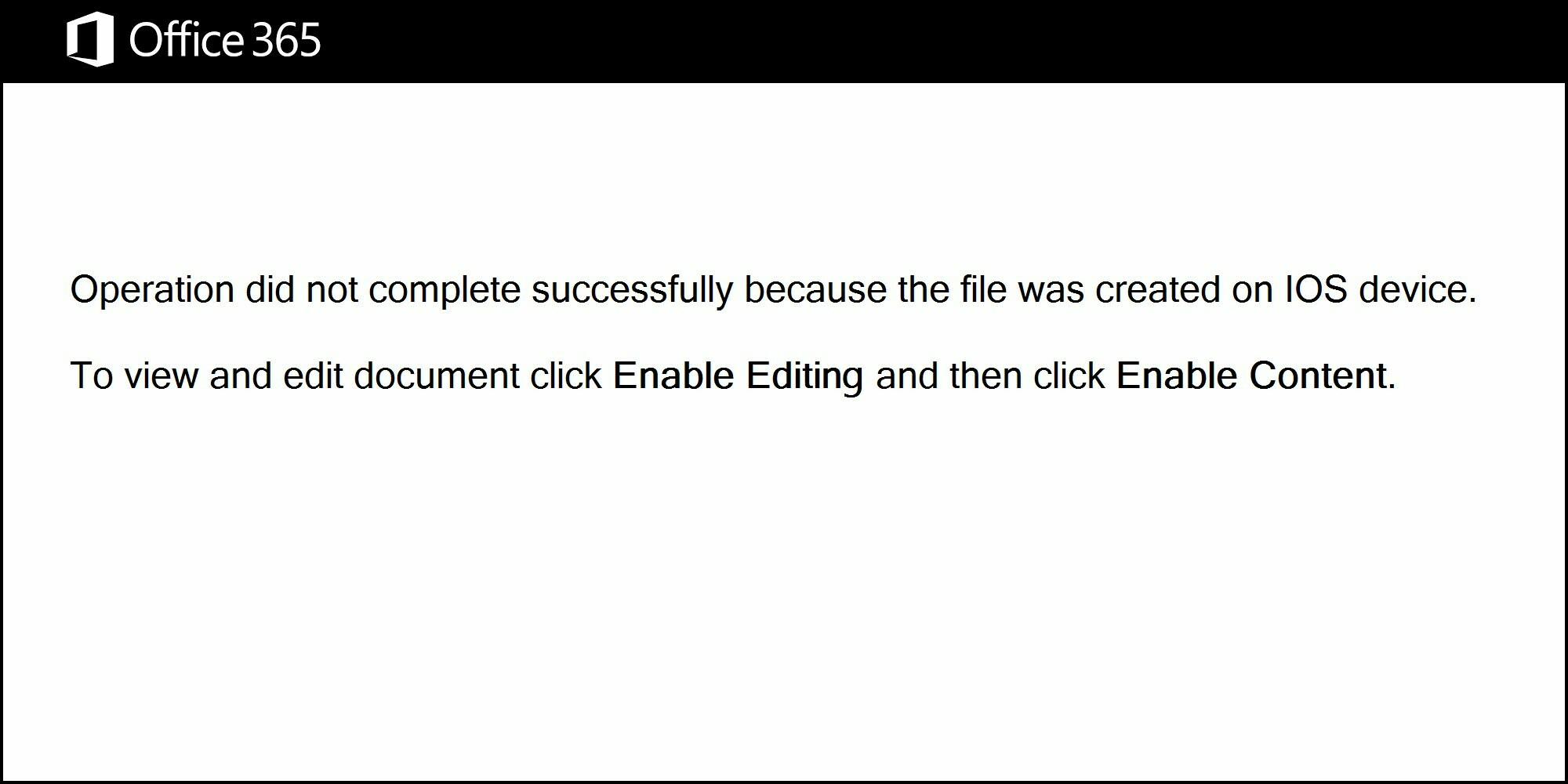}
\includegraphics[width=0.45\textwidth]{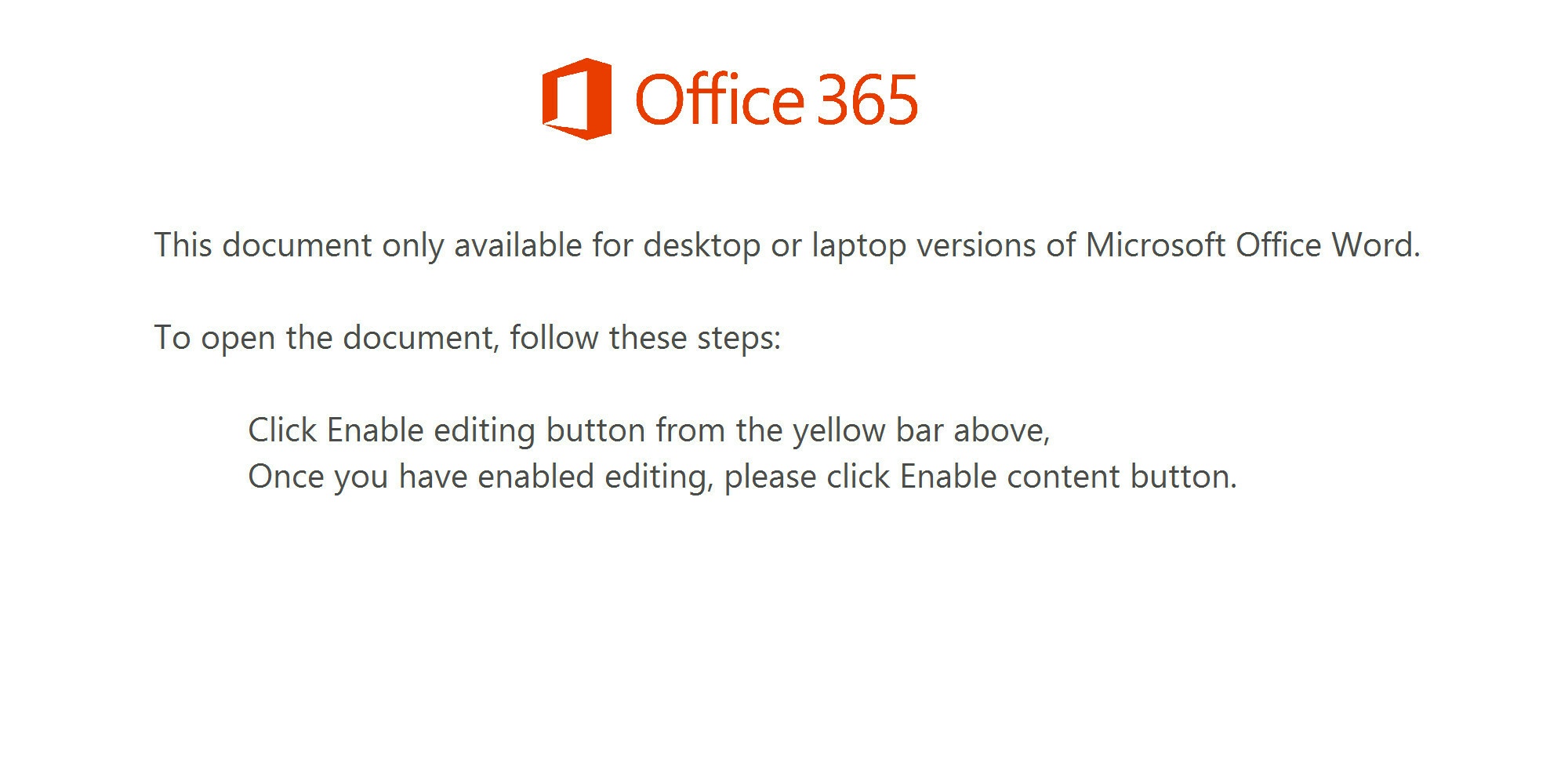}
\includegraphics[width=0.45\textwidth]{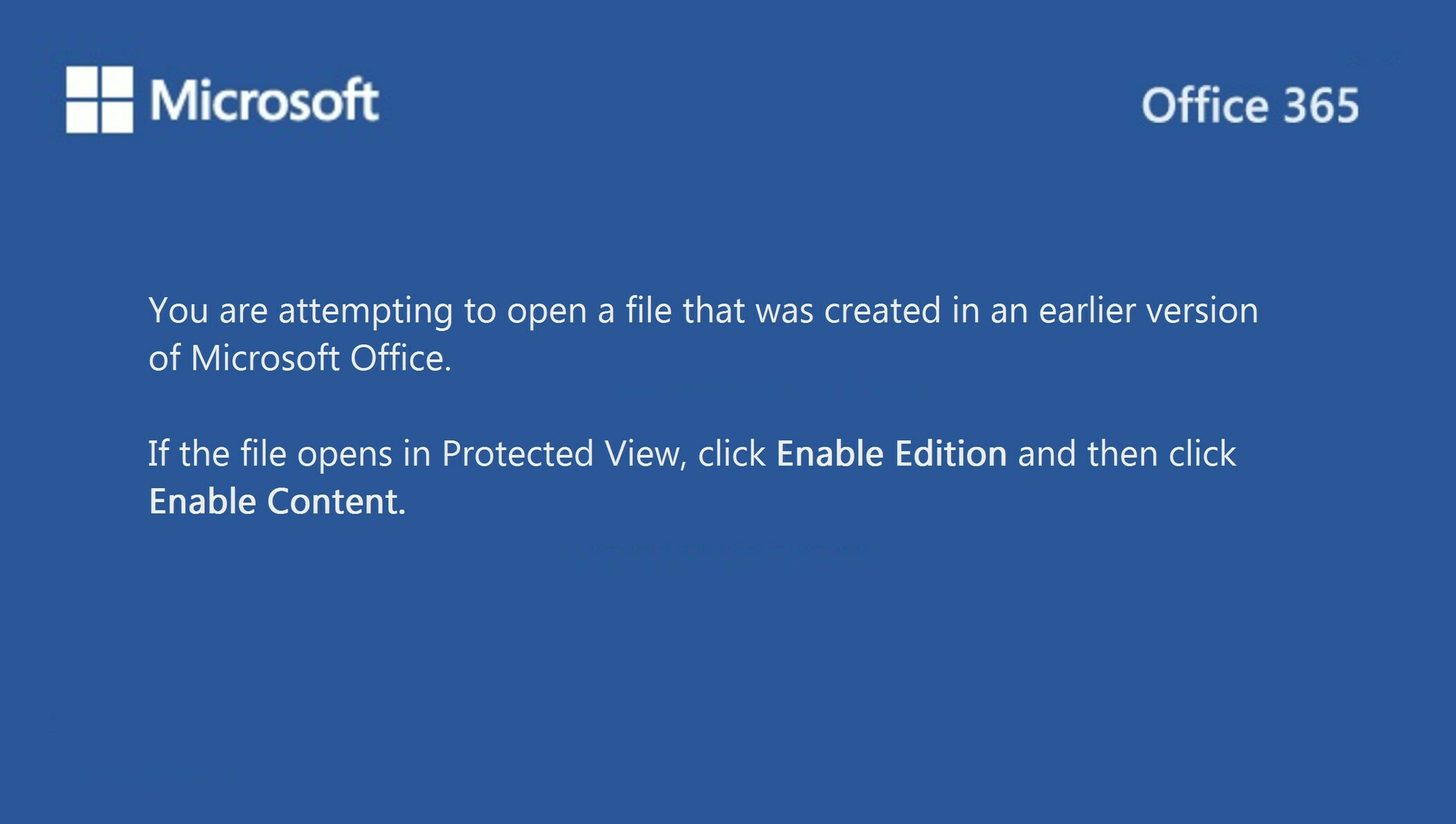}
\includegraphics[width=0.45\textwidth]{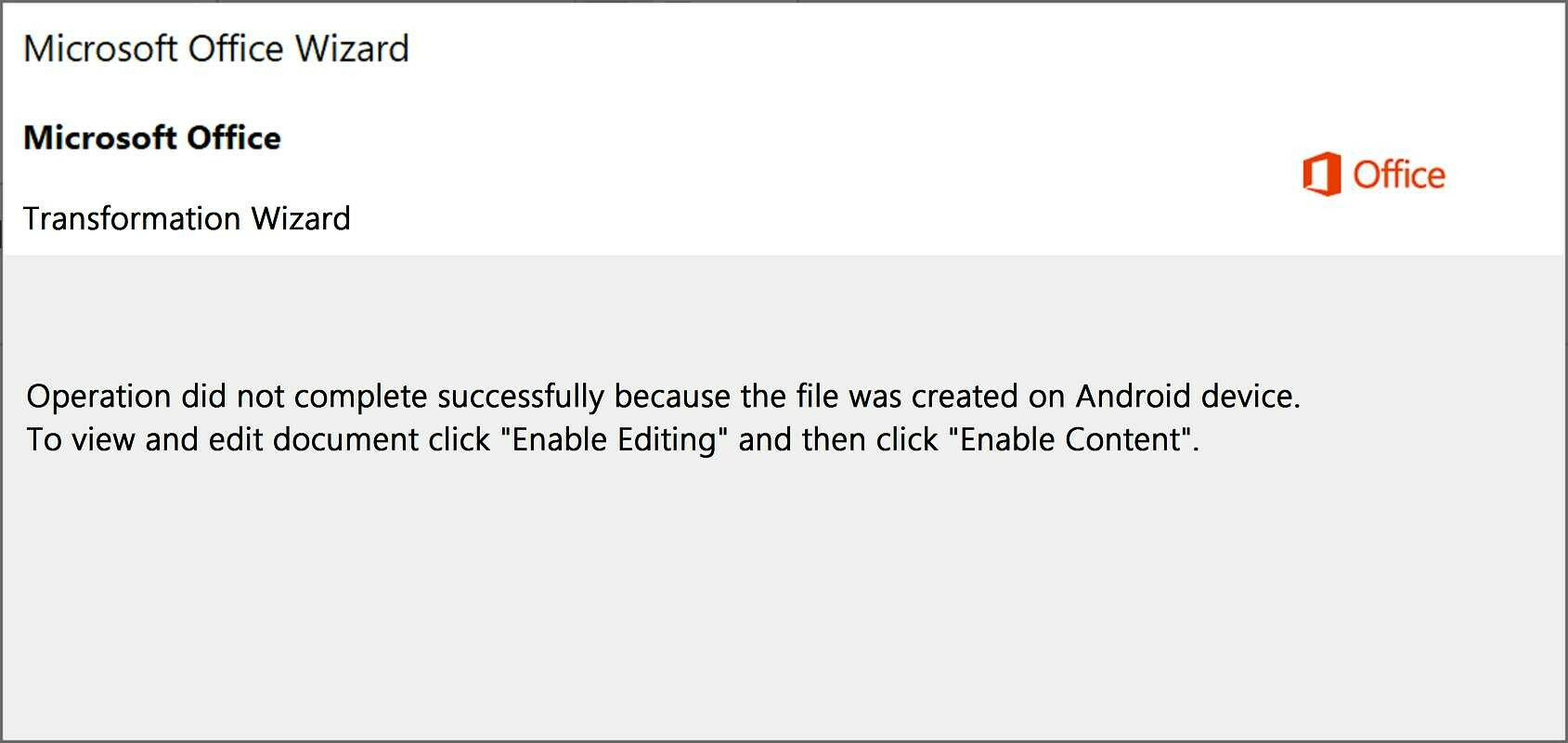}
\includegraphics[width=0.45\textwidth]{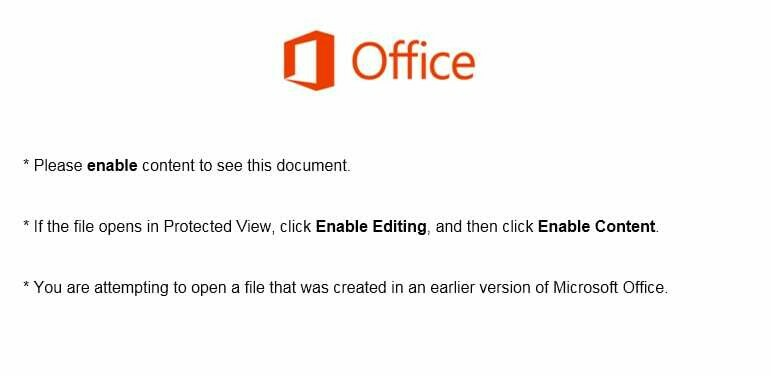}
    \caption{Templates of images displayed on Word.}
    \label{fig:templates}
\end{figure}

\begin{figure}[!ht]
    \centering
\includegraphics[width=0.45\textwidth]{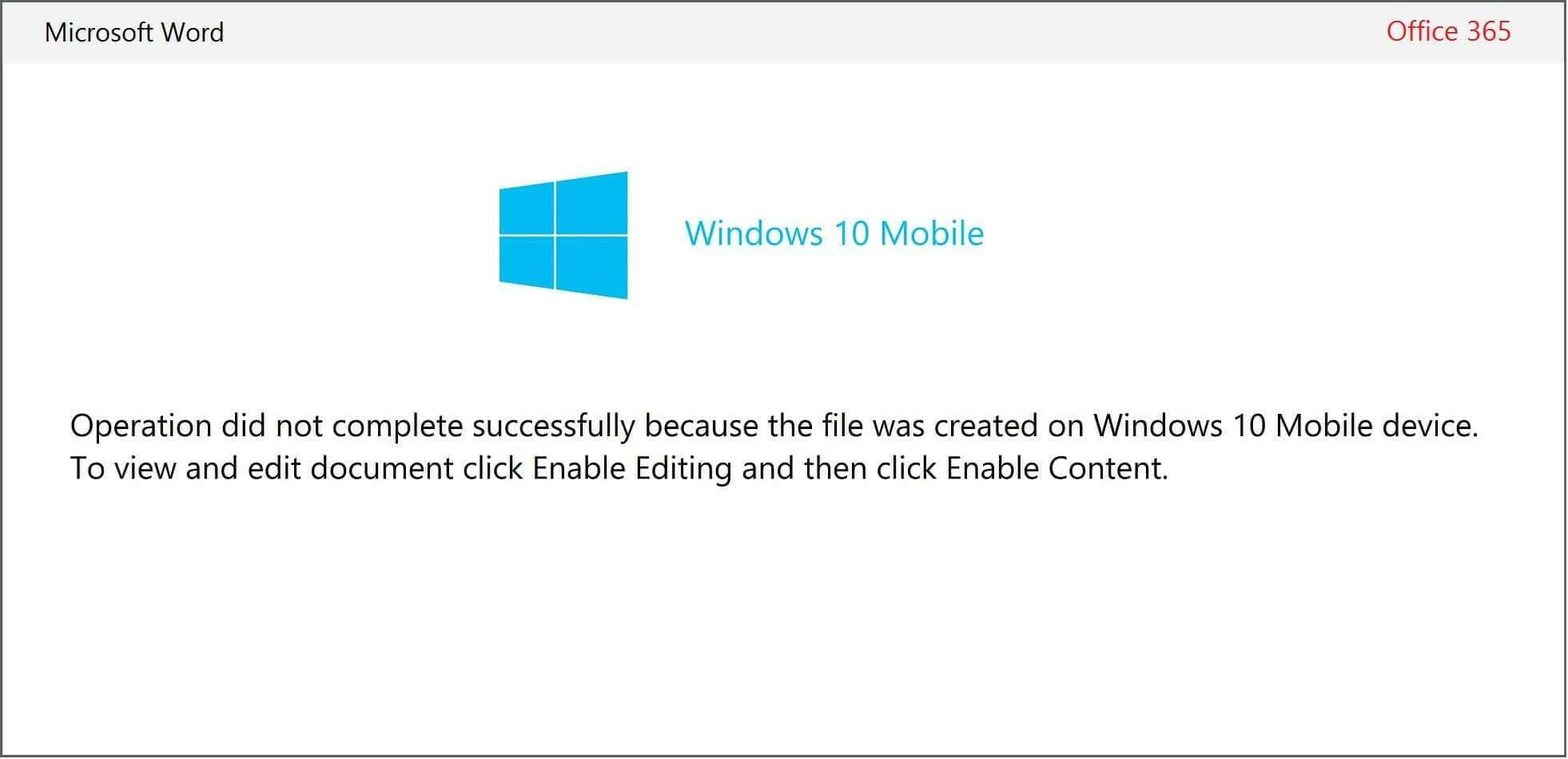}
\includegraphics[width=0.45\textwidth]{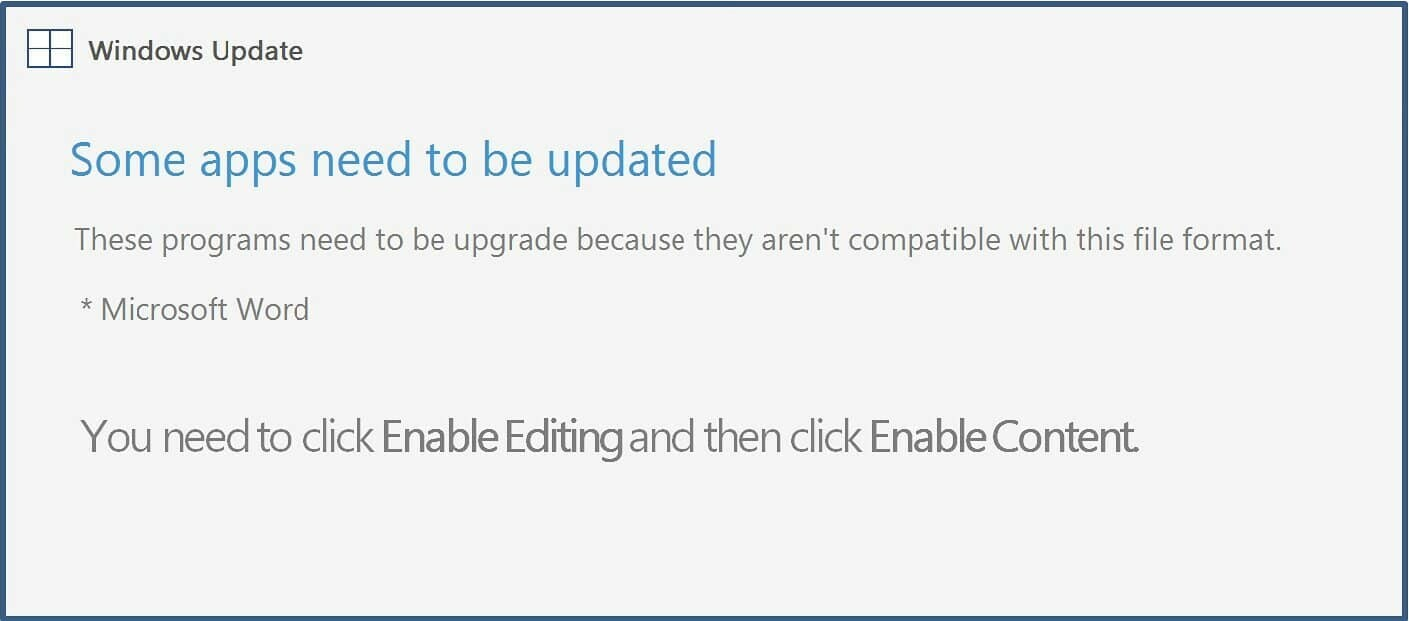}
\includegraphics[width=0.45\textwidth]{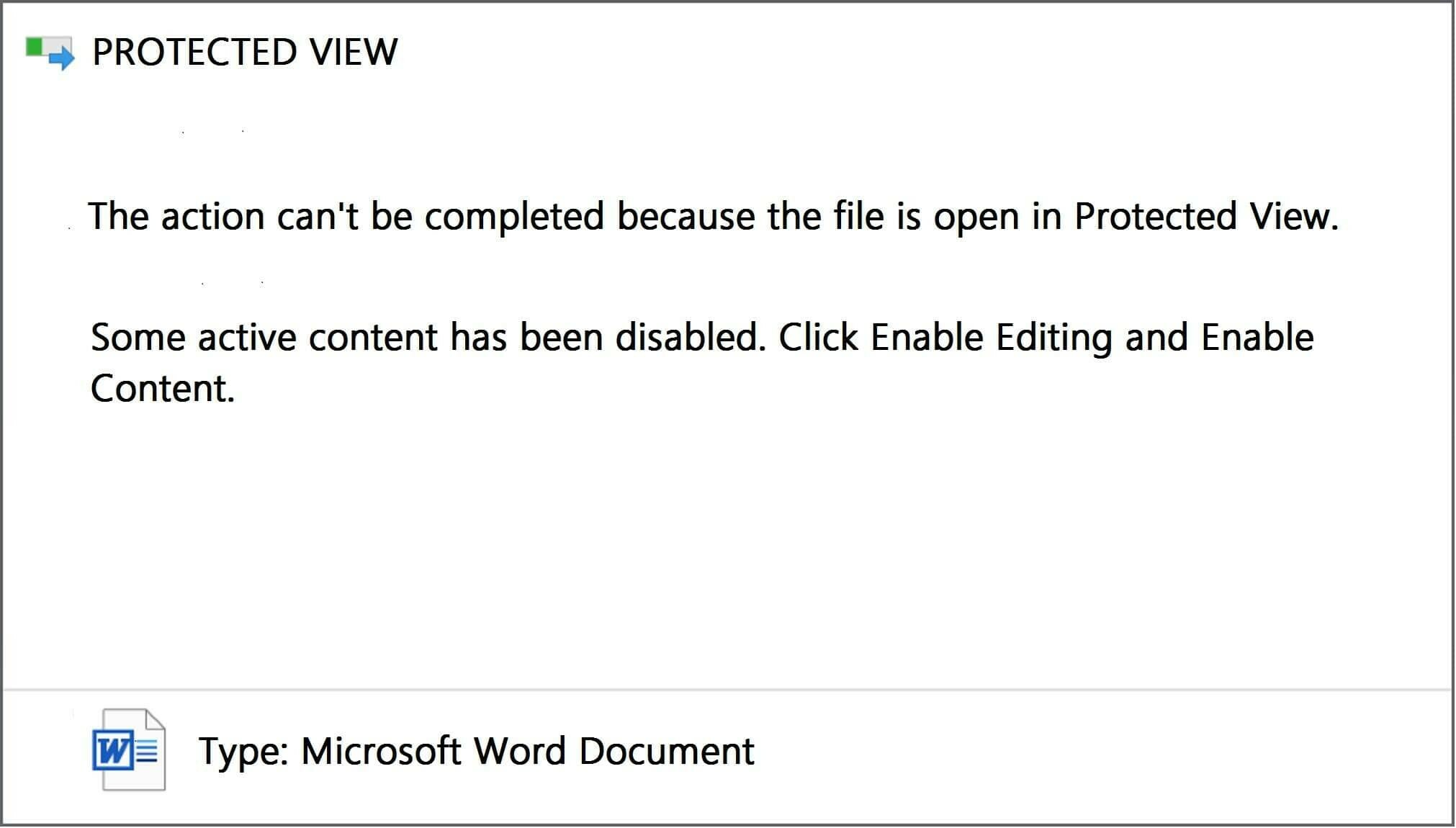}
\includegraphics[width=0.45\textwidth]{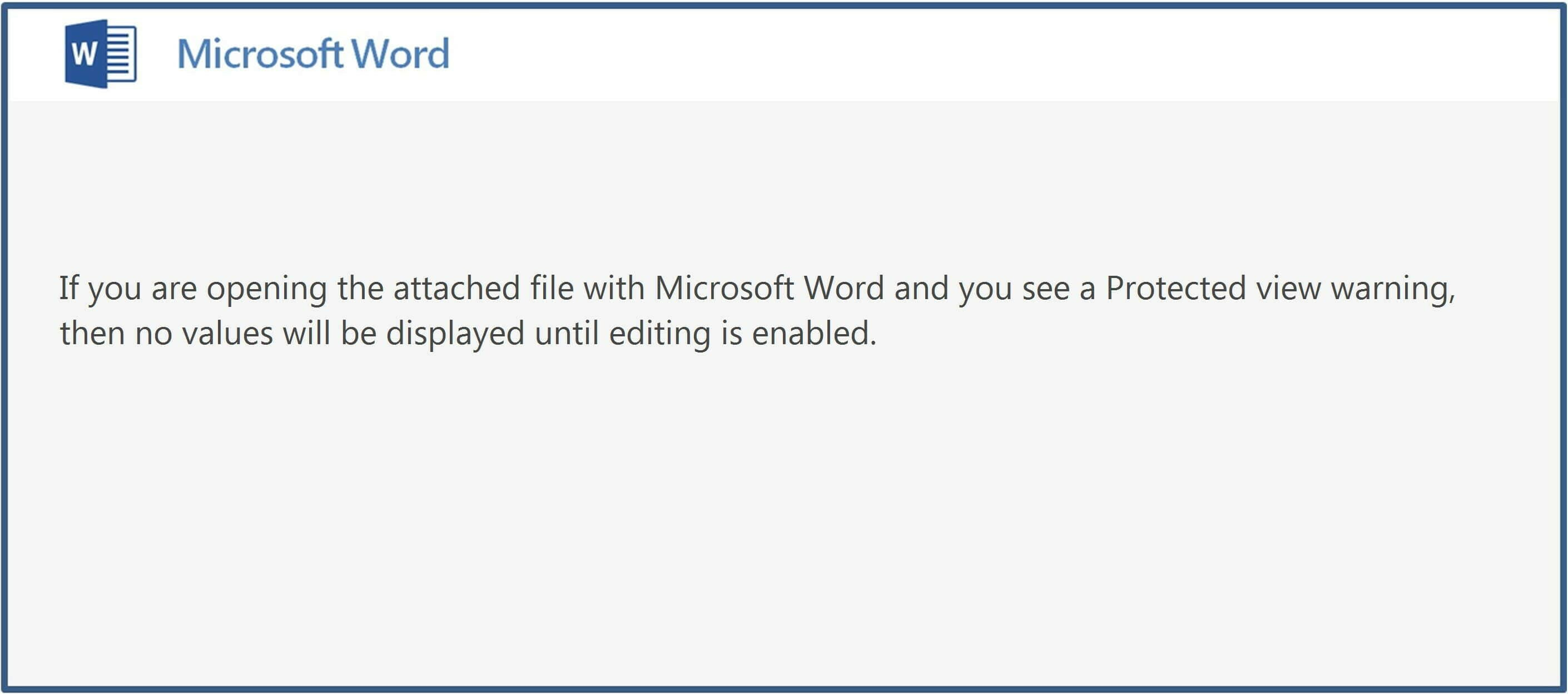}
\includegraphics[width=0.45\textwidth]{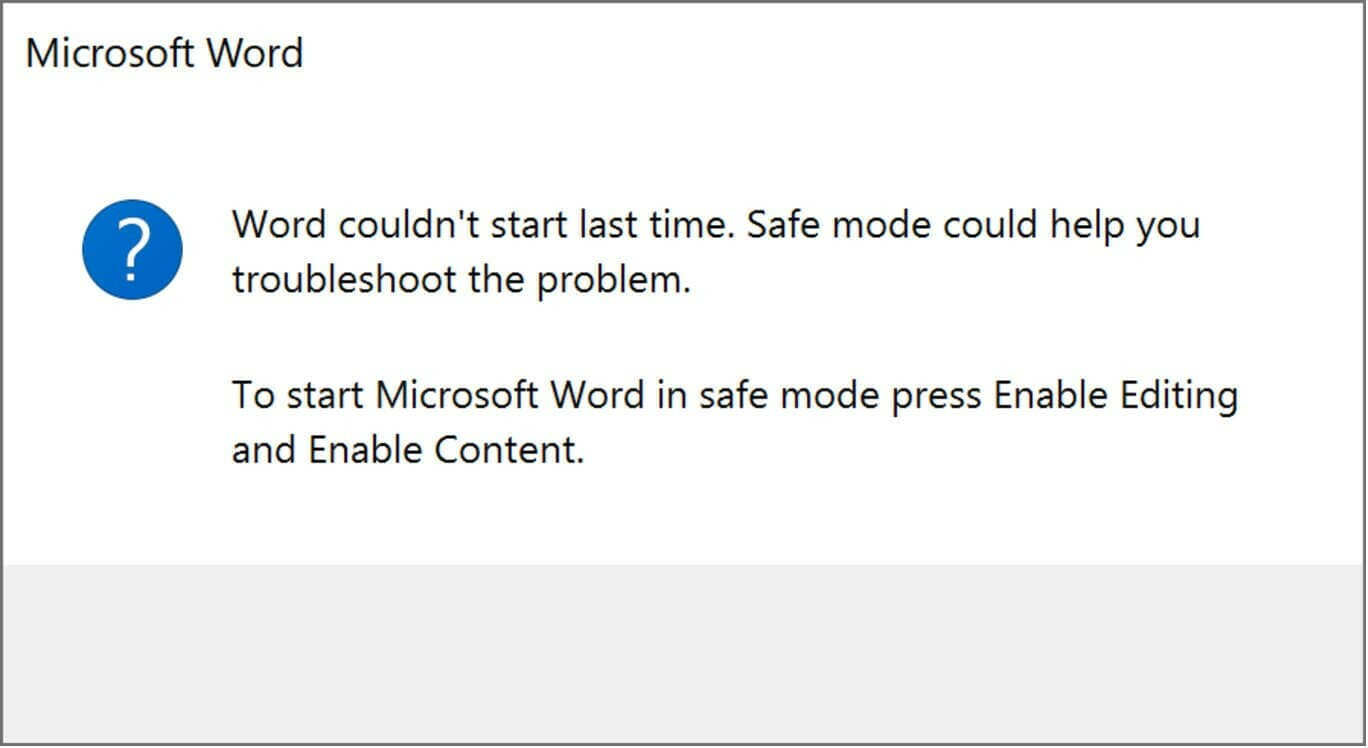}
\includegraphics[width=0.45\textwidth]{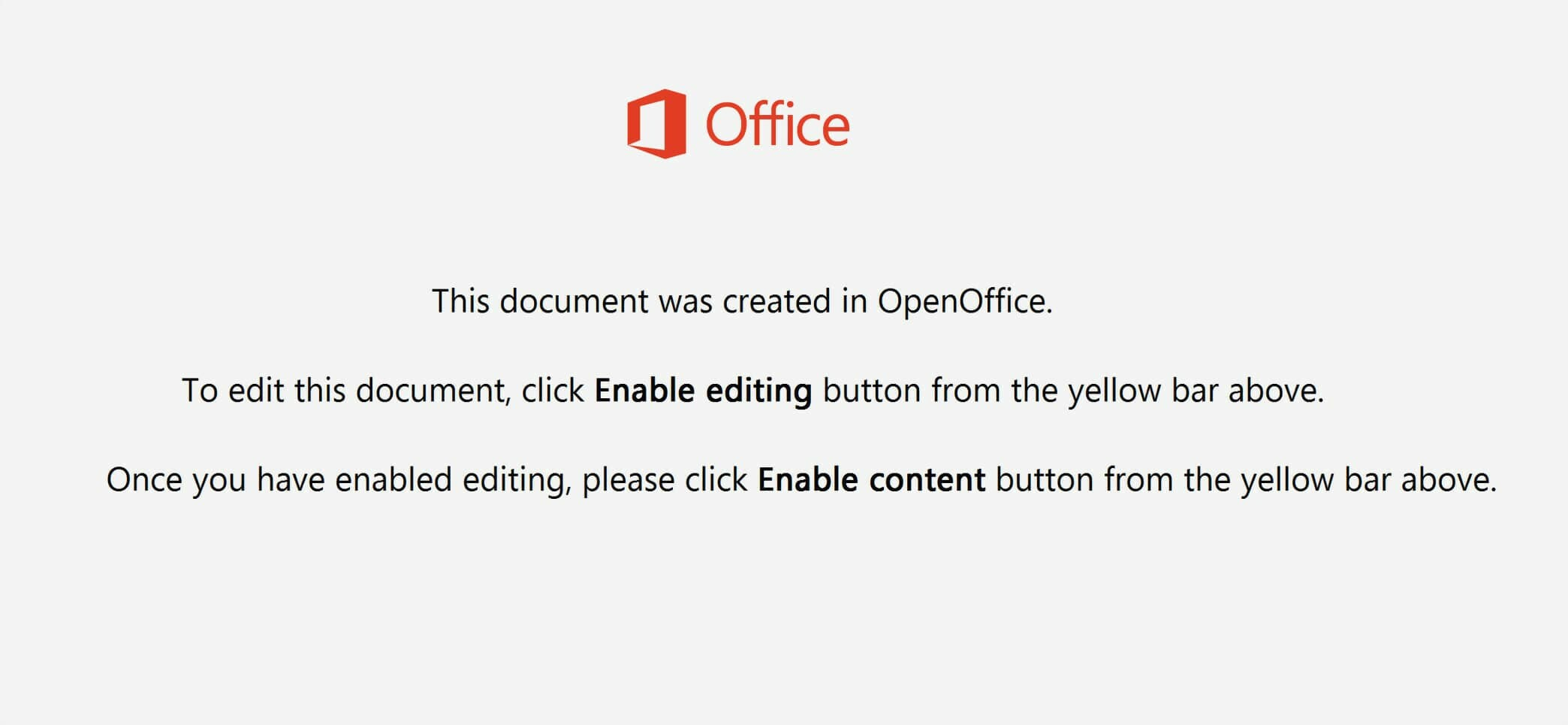}
\includegraphics[width=0.45\textwidth]{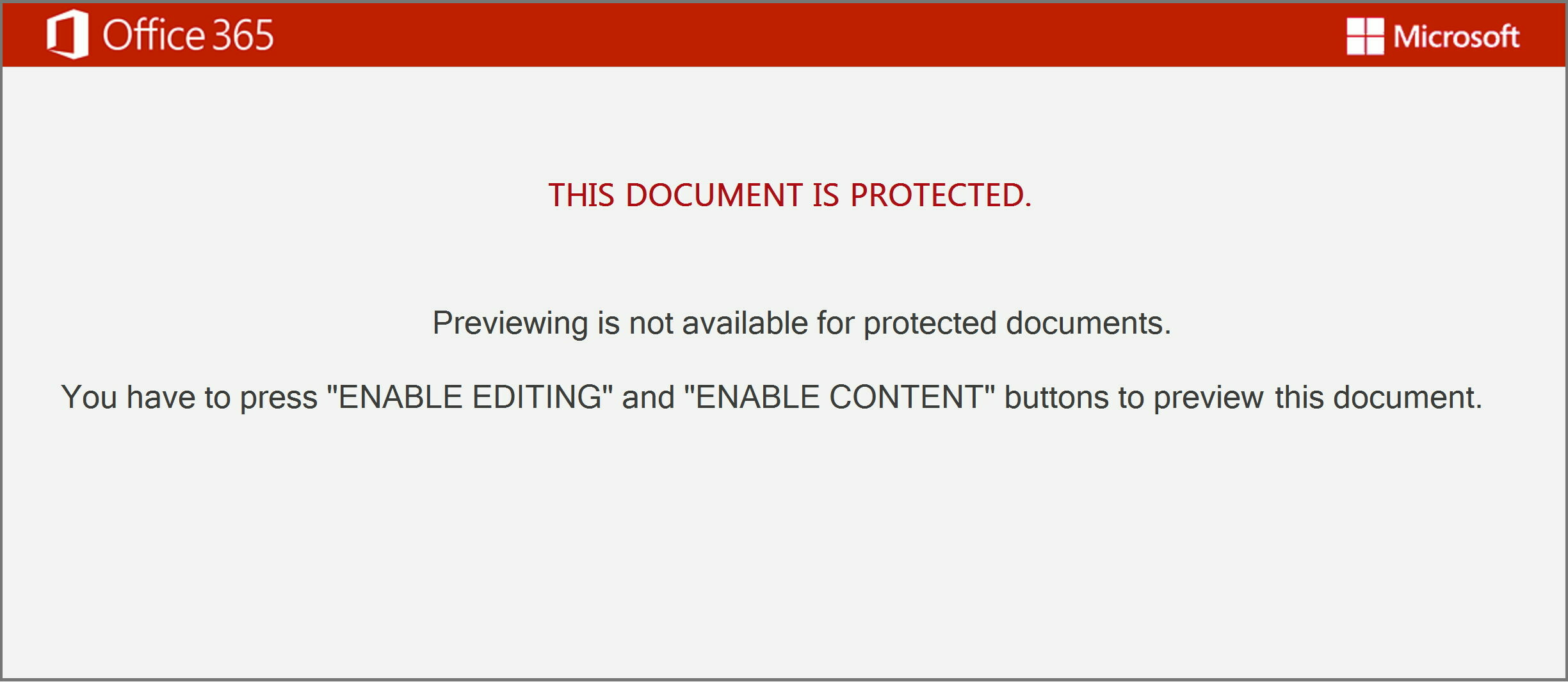}
\includegraphics[width=0.45\textwidth]{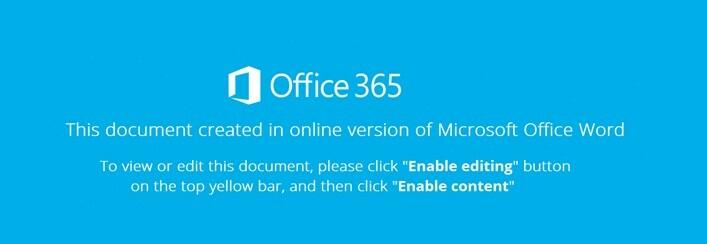}
\includegraphics[width=0.45\textwidth]{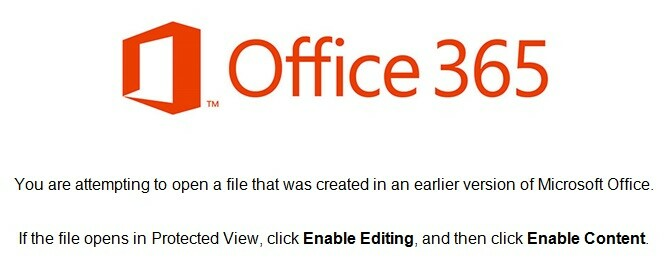}
\includegraphics[width=0.45\textwidth]{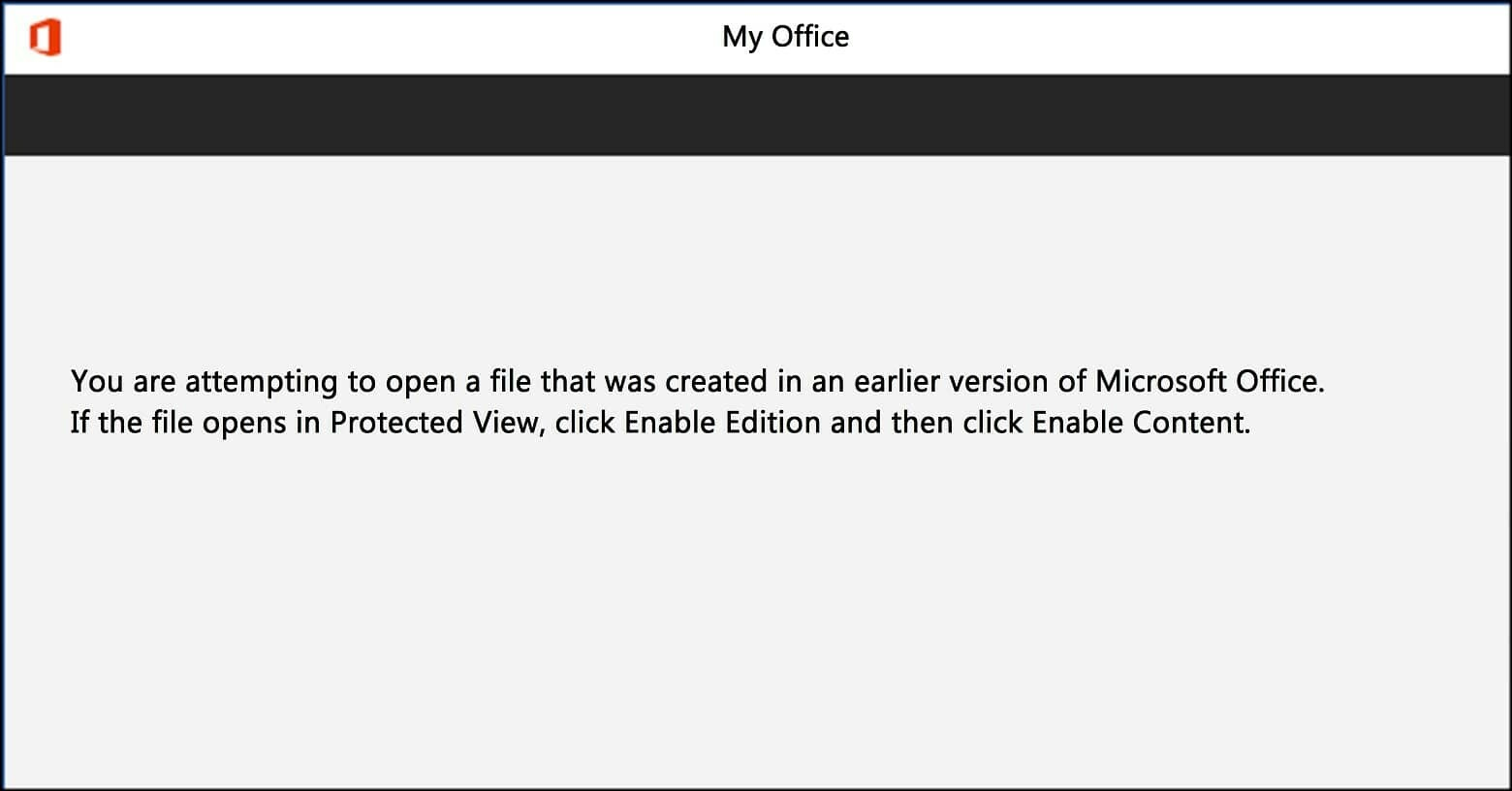}
\caption{Templates of images displayed on Word, continued.}
\label{fig:templates2}
\end{figure}

In general, we have observed 18 versions of the malicious Word files, with minor main changes identified in the obfuscation of the VBA code and the final Powershell payload. In all cases, the adversary tries to lure the victim that the document is valid and some technical issue, due to, e.g. compatibility prevents it from being loaded. In this context, the victim is prompted to, e.g. \textit{Enable Content}, so that the document is correctly loaded. In doing so, the victim allows the execution of the malicious VBA code, which is contained in the document (of course enabling content/macros is needed only if this is not allowed by default). To trick the victim into running the malicious VBA code, a specially crafted image is displayed to the user, which informs the victim of the ``Document loading'' technical issue. Several different image templates were observed to be used, see Figures \ref{fig:templates} and \ref{fig:templates2}.

\begin{figure}[!ht]
    \centering
    \includegraphics[width=\textwidth]{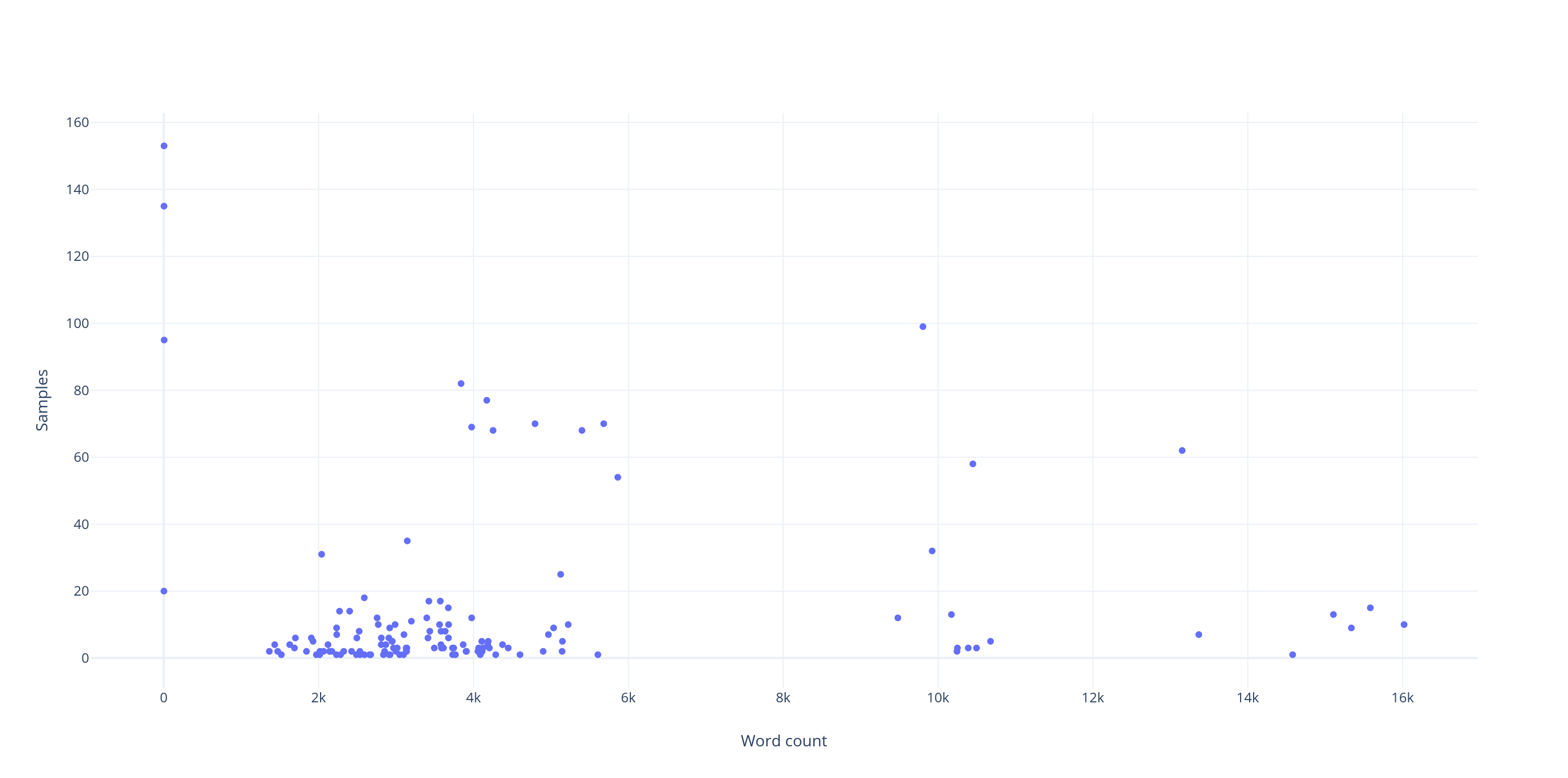}
    \caption{Samples with same word count.}
    \label{fig:word_count}
\end{figure}

Interestingly, the reported total edit time in all samples is 0, the revision number is 1, and the creation time coincides with the last saved time (in few occurrences, due to the minute change there is a diff of 1 minute), so the reported edit time is 0. The bulk of documents (1747) are one page long while 137 are two pages long, see Figure \ref{fig:word_count}.

\begin{figure}[!ht]
    \centering
    \includegraphics[width=\textwidth]{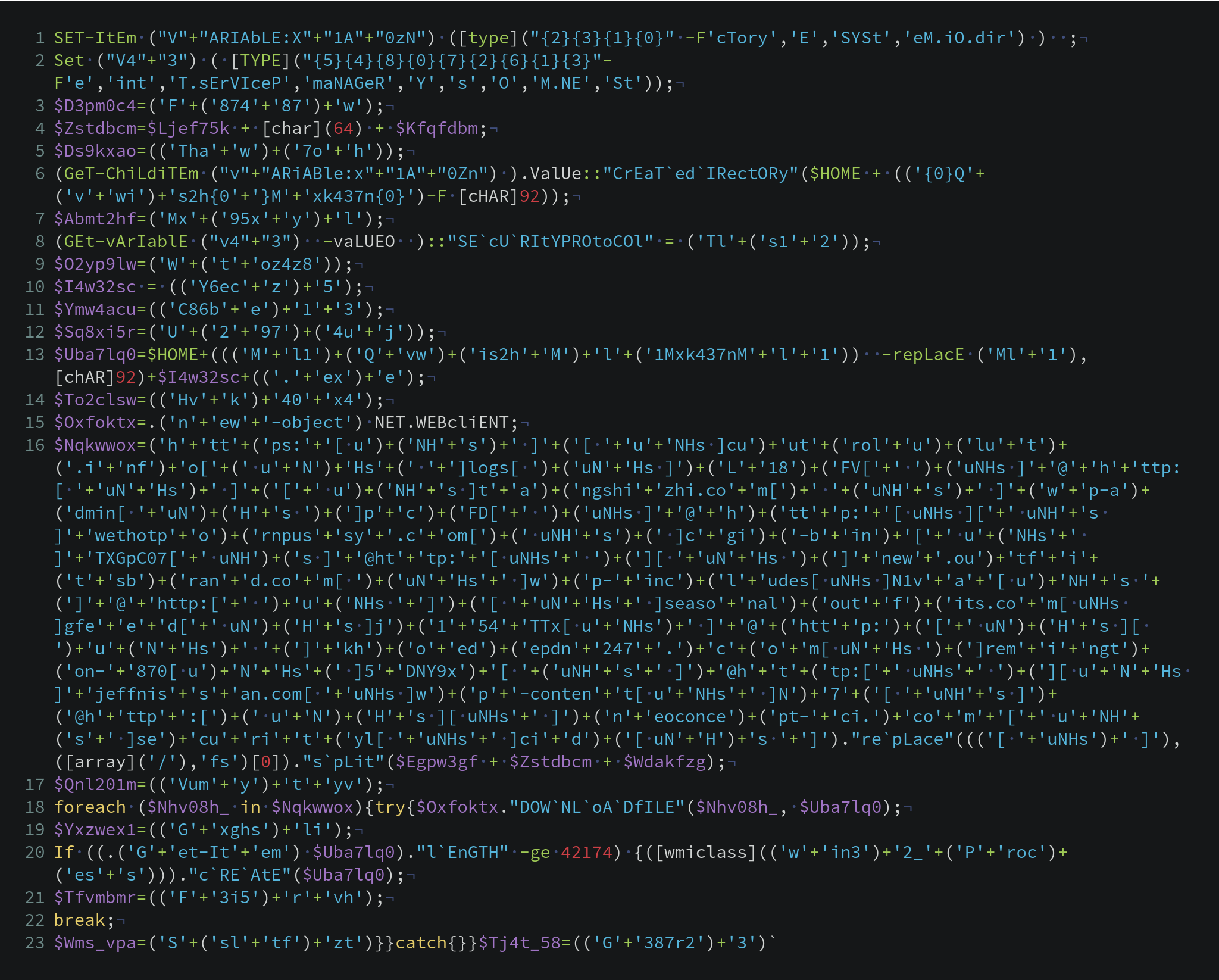}
    \caption{An obfuscated PowerShell payload.}
\label{lst:obf_ps}
\end{figure}

\begin{figure}[!ht]
    \centering
    \includegraphics[width=\textwidth]{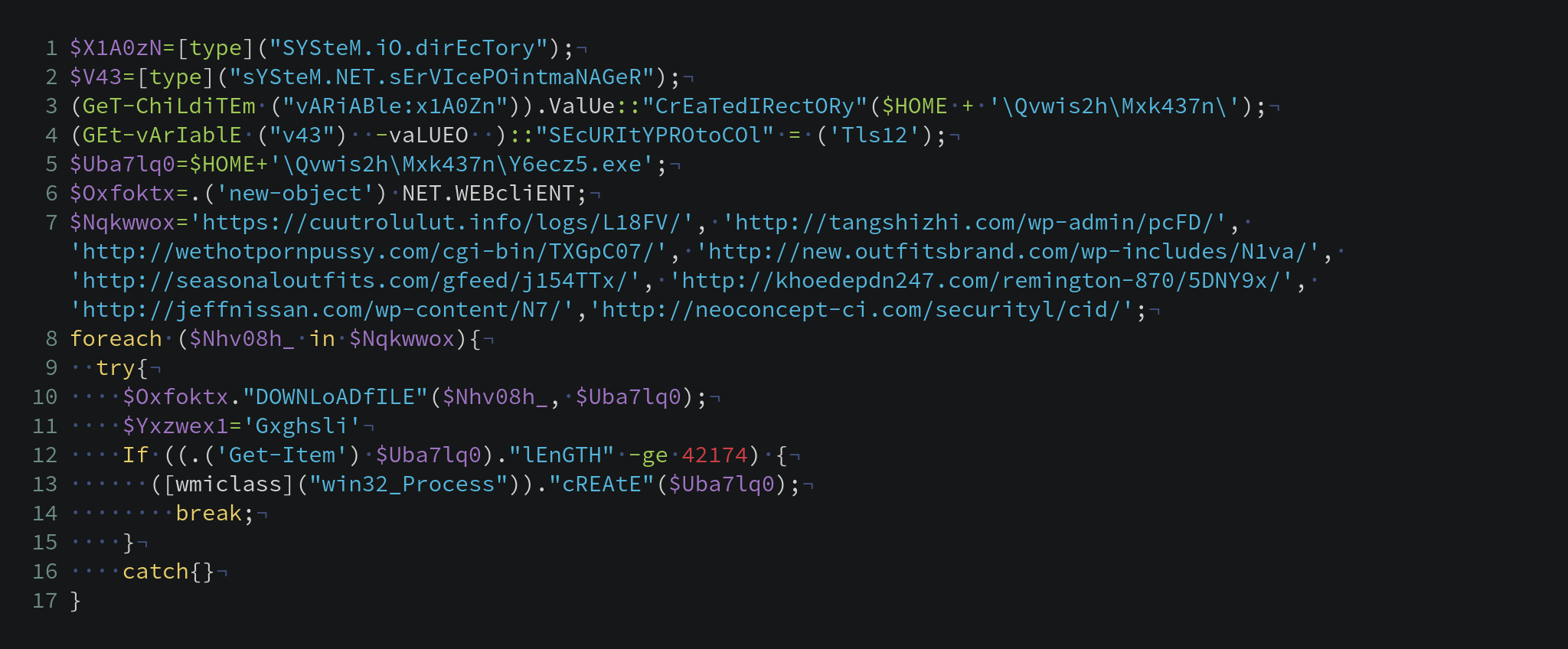}
    \caption{Deobfuscated PowerShell payload of Figure \ref{lst:deobf_ps}.}
\label{lst:deobf_ps}
\end{figure}

The PowerShell payload comes in the form of a base64 encoded string, which is decoded to execute an obfuscated Powershell script that tries to download a malicious binary of Emotet from various URLs. Once the Powershell script manages to download the binary, the latter is being executed by the script. A decoded version of an obfuscated PowerShell script, which was obtained from one of the analysed Microsoft Word files, is illustrated in Figure \ref{lst:obf_ps}, while its deobfuscated form is illustrated in Figure \ref{lst:deobf_ps}. Based on the code characteristics of the obfuscated PowerShell code, most likely, the tool used to produce the Powershell code is Daniel Bohannon's ``\textit{Invoke-Obfuscation}''\footnote{\url{https://github.com/danielbohannon/Invoke-Obfuscation}} \cite{bohannon2017revoke}.

\begin{figure}[!ht]
    \centering
    \includegraphics[width=\textwidth]{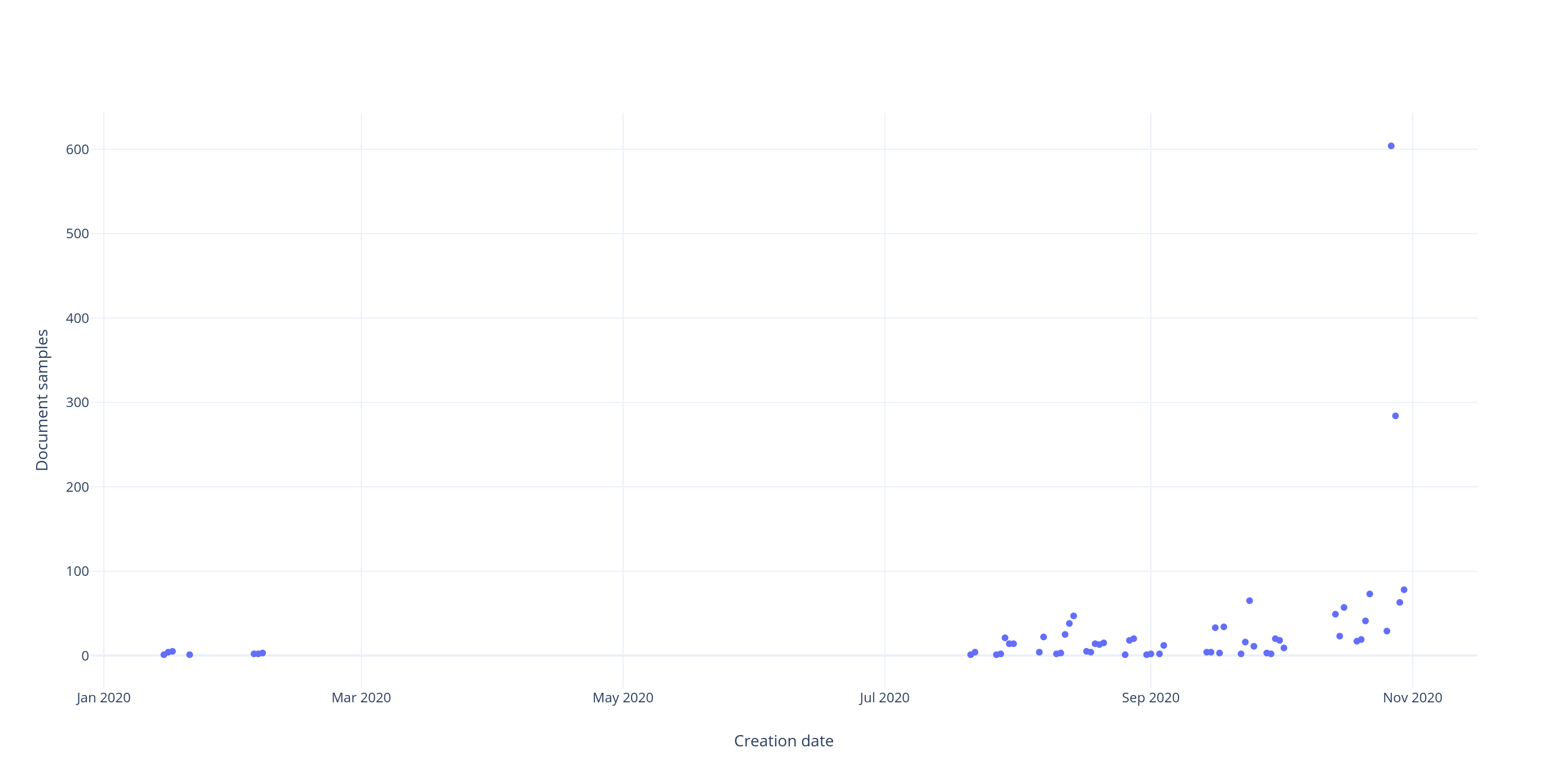}
    \caption{Documents per day in the dataset.}
    \label{fig:docs_per_day}
\end{figure}

\section{Statistics from  downloaded Emotet binaries}

Many samples try to get screenshots and hook the victim's keyboard to record keystrokes. Several samples try to perform privilege escalation through \texttt{Advapi32.dll}. VM detection (e.g. Qemu, Virtual Box, VMWare) and detection of sandbox environment (Joe sandbox) are also performed by many of these samples. Additionally, checks for debuggers through Structured Exception Handling is performed in almost all samples. Furthermore, the detection of the existence of Frida and sysmon in processes are also performed by many samples.

The capabilities of the downloaded binaries according to the MITRE ATT\&CK technique classification \cite{attack} are summarised in Table \ref{tbl:attack}.

\begin{table}[!th]
    \centering
    \begin{tabular}{lp{3.75in}r}
    \toprule
    \textbf{Vector} & \textbf{Technique} & \textbf{Samples}\\
    \midrule
Execution & Shared Modules [T1129] & 741\\
\rowcolor{red!25}Persistence & Boot or Logon Autostart Execution: Registry Run Keys / Startup Folder [T1547.001] & 7\\
Privilege Escalation & Access Token Manipulation [T1134] & 34\\
\rowcolor{red!25} & Obfuscated Files or Information [T1027] & 652\\
\rowcolor{red!25} & Process Injection [T1055] & 380\\
\rowcolor{red!25} & Virtualization/Sandbox Evasion:System Checks [T1497.001] & 352\\
\rowcolor{red!25} & Indicator Removal on Host:Timestomp [T1070.006] & 9\\
\rowcolor{red!25}\multirow{-5}{*}{Defense Evasion} & Virtualization/Sandbox Evasion:User Activity Based Checks [T1497.002] & 6\\
Credential Access & Data from Local System [T1005] & 7\\

\rowcolor{red!25} & Query Registry [T1012] & 367\\
\rowcolor{red!25} & System Information Discovery [T1082] & 359\\
\rowcolor{red!25} & File and Directory Discovery [T1083] & 335\\
\rowcolor{red!25} & Application Window Discovery [T1010] & 124\\
\rowcolor{red!25} & Process Discovery [T1057] & 41\\
\rowcolor{red!25} & System Owner/User Discovery [T1033] & 31\\
\rowcolor{red!25} \multirow{-7}{*}{Discovery}& System Service Discovery [T1007] & 3\\

\multirow{3}{*}{Collection} & Input Capture:Keylogging [T1056.001] & 262\\
 & Screen Capture [T1113] & 13\\
 & Clipboard Data [T1115] & 4\\
\bottomrule
    \end{tabular}
    \caption{Capabilities of downloaded by Emotet executables according to MITRE ATT\&CK techique classification.}
    \label{tbl:attack}
\end{table}

\begin{figure}[!th]
    \centering
    \includegraphics[width=\textwidth]{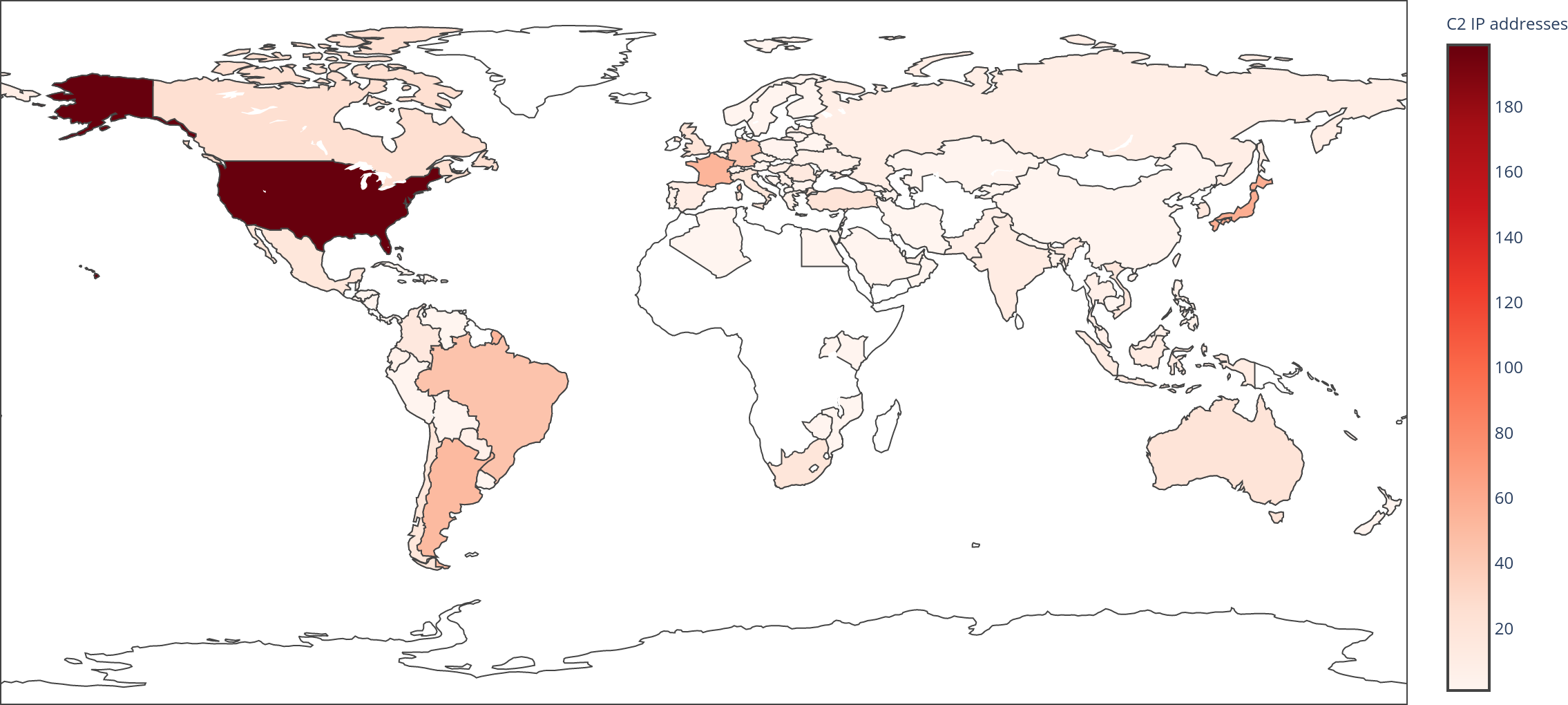}
    \caption{Geolocation of C2 servers}
    \label{fig:c2}
\end{figure}

The executables try to connect with C2 servers of Emotet, which are hardcoded in the binary. Our analysis extracted 1054 unique IP addresses whose geolocation is illustrated in Figure \ref{fig:c2}. It is worth noting that the C2 servers are not always available as they follow a round-robin strategy, to avoid being flagged as malicious \cite{disposable}.

\begin{figure}[!th]
    \centering
    \includegraphics[width=\textwidth]{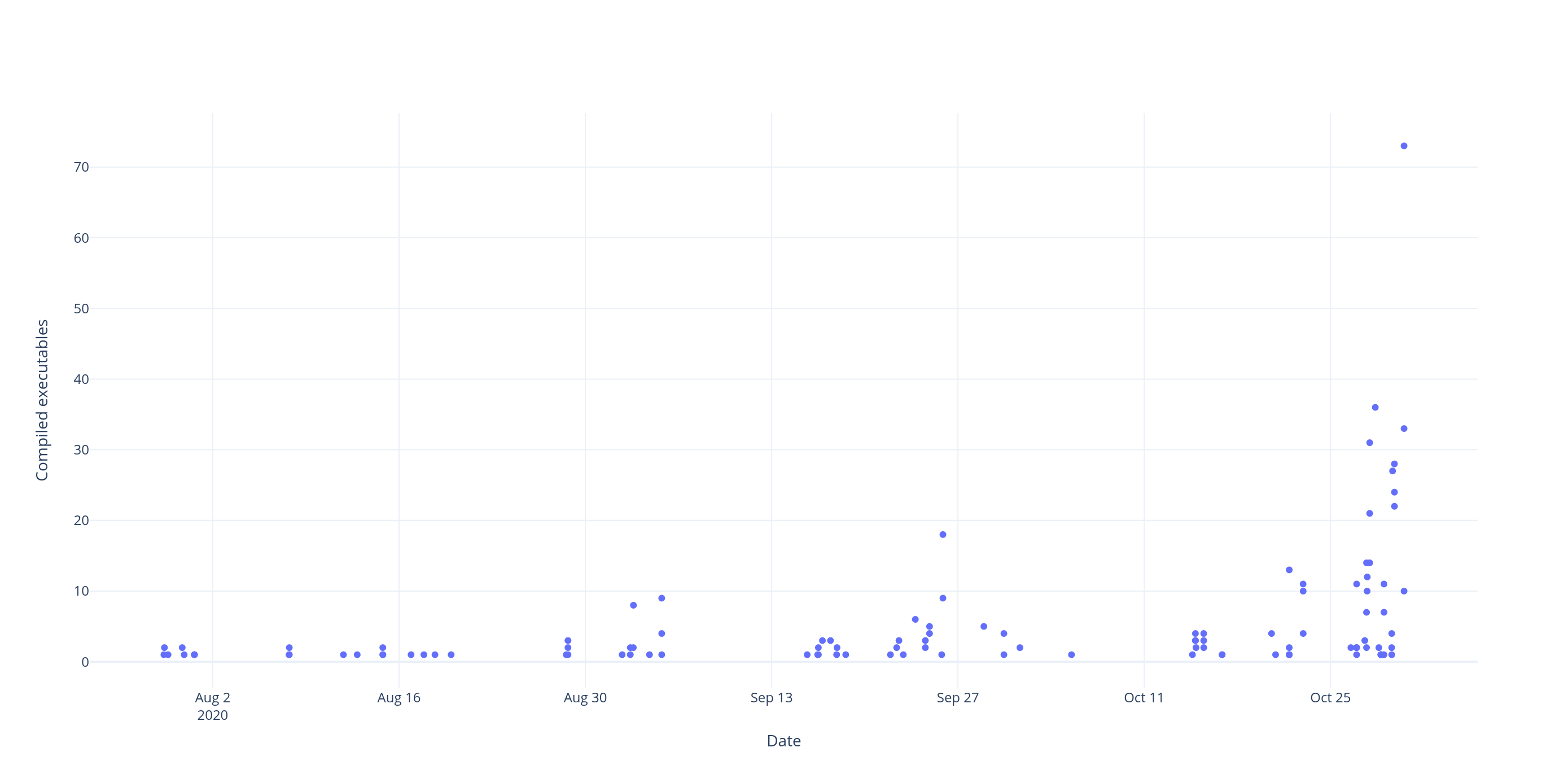}
    \caption{Compilation time per date.}
    \label{fig:compiled_by_date}
\end{figure}

\begin{figure}[!th]
    \centering
    \includegraphics[width=\textwidth]{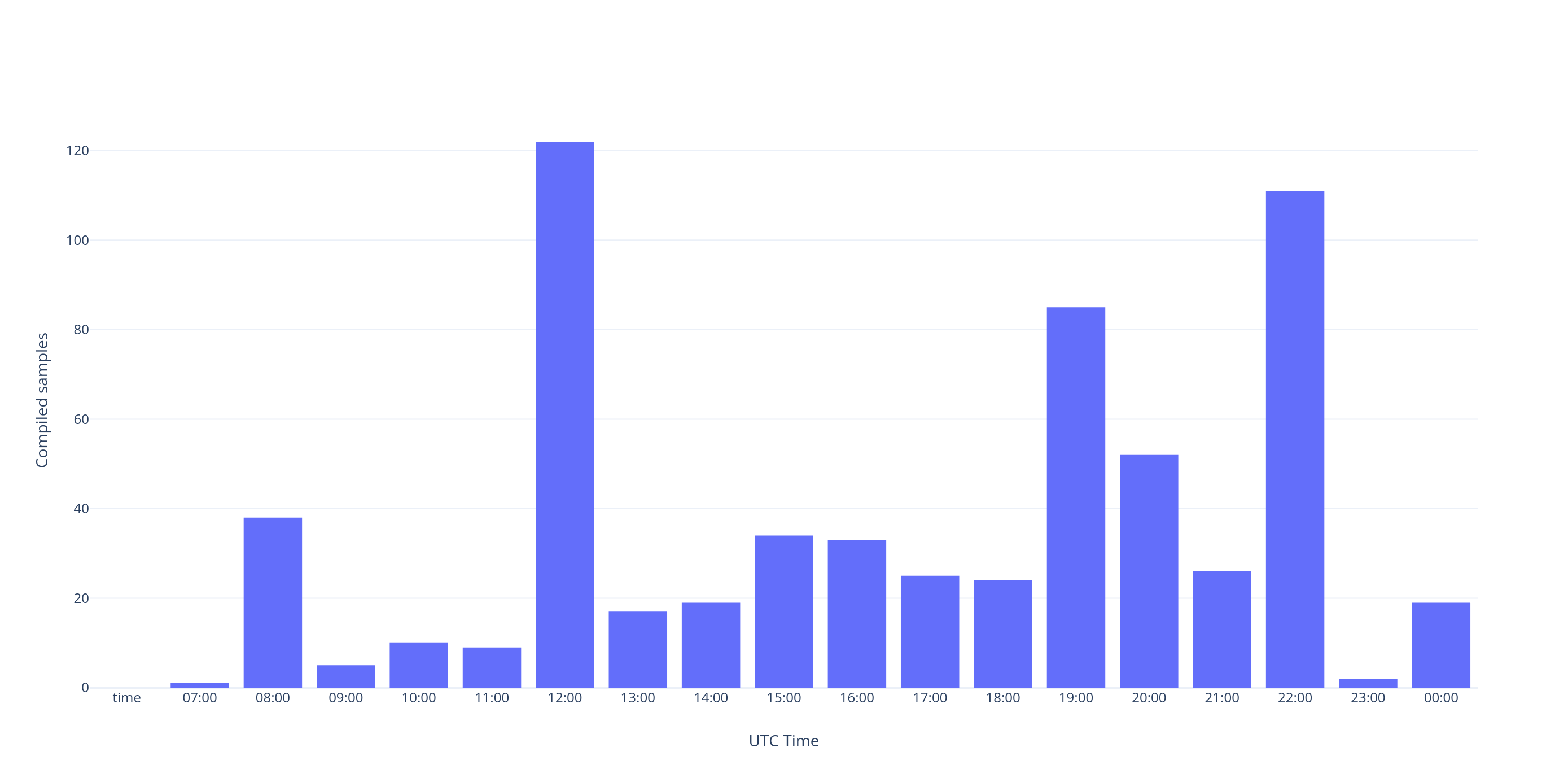}
    \caption{Compilation times of samples.}
    \label{fig:compiled_by_time}
\end{figure}

\begin{figure}[!th]
    \centering
    \includegraphics[width=\textwidth]{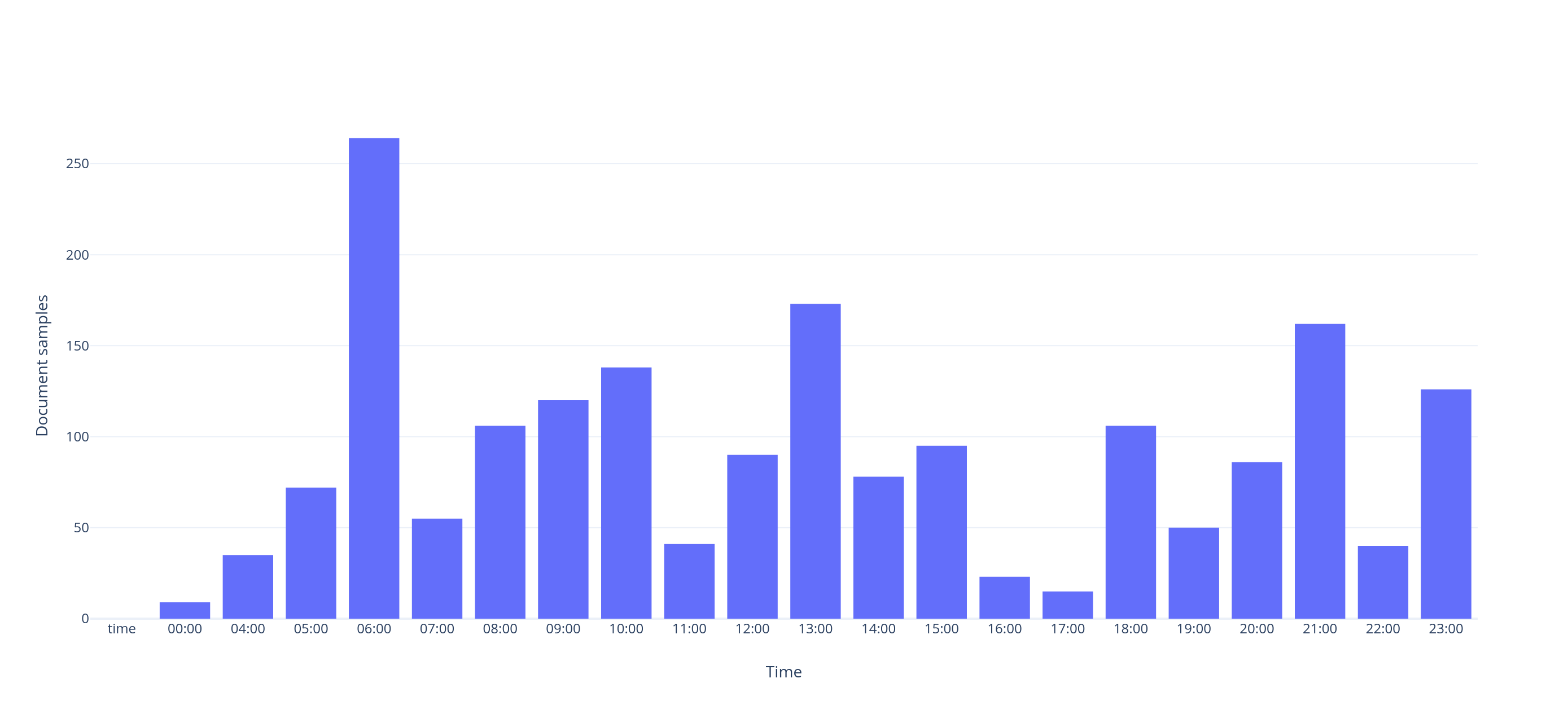}
    \caption{Documents per creation time in the dataset.}
    \label{fig:docs_per_time}
\end{figure}

The spread of samples according to the reported compilation time is illustrated in Figure \ref{fig:compiled_by_date}. Moreover, Figure \ref{fig:compiled_by_time} illustrates a pattern in the reported  time, indicating that the ``working'' hours span from 8AM to 00:00AM. Notably, the pattern for documents differentiates as it spans from 4:00 AM to 00:00AM, see Figure \ref{fig:docs_per_time}.

\begin{figure}[!th]
    \centering
    \includegraphics[width=.75\textwidth]{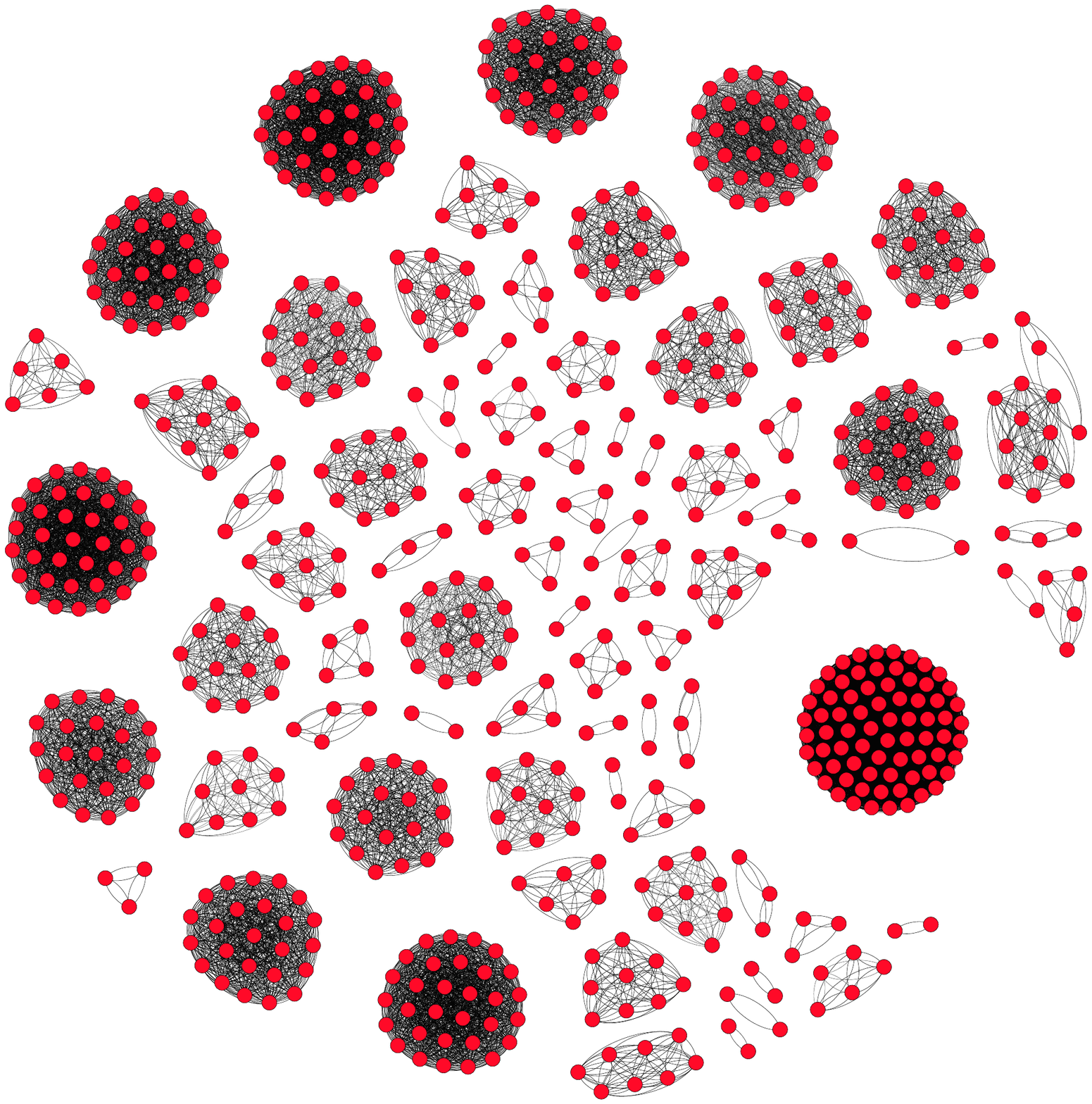}
    \caption{SSDeep clustering}
    \label{fig:ssdeep}
\end{figure}

The samples that are used are mainly compiled in Visual Studio by utilizing various versions of Visual Basic and C++. The packer that is mainly used is Armadillo; however, there are several versions of the packer that are used. The latter is verified by the reuse of some base code from several open-source projects to masquerade their binaries and impede their analysis and attribution such as the ones mentioned below:
\begin{itemize}
\item \url{https://www.codeproject.com/Articles/2687/Chat-With-US-DI}
\item \url{https://www.codebus.net/d-iAK.html}
\end{itemize}

In general, the samples are clustered in distinct and well-formed clusters that are confirmed by the produced, during the conducted analysis, SSDeep clusters (cf. Figure \ref{fig:ssdeep}) \cite{191669}, as well as through Imphash \cite{naik2019cyberthreat} and RichPE metadata hash \cite{webster2017finding,dubyk2019sans}, thus indicating similar patterns in terms of execution and impact.

\begin{table}[!th]
\centering
\begin{tabular}{lr|lr}
\hline
\textbf{Samples} & \textbf{Imphash} & \textbf{Samples} & \textbf{RichPE metadata hash} \\ \hline
116              & c9f7e...         & 116              & 199a6...         \\
74               & 50f8a...         & 74               & c5df6...         \\
66               & a1ffb...         & 66               & 7026a...         \\ 
44               & 949a5...         & 45               & 84ee2...         \\
34               & ee32a...         & 36               & 1de9f...         \\
27               & 6a92a...         & 27               & 5c985...         \\
27               & 521d2...         & 27               & 3e73c...         \\
23               & 875a1...         & 25               & 20f69...         \\
19               & 44be8...         & 19               & 7d848...         \\
14               & 5da88...         & 16               & 9655a...         \\
14               & 195da...         & 15               & 46c23...         \\
12               & 6f669...         & 14               & 8e0a5...         \\
10               & ead6f...         & 12               & 575fd...         \\
10               & bc97b...         & 10               & 4d0fe...         \\
9                & cbf12...         & 10               & 3de2a...         \\ \hline
\end{tabular}
\caption{Top 15 Imphash and RichPE metadata hash clusters.}
\label{tbl:top_clusters}
\end{table}

From the PDB remnants, the user names that are identified are BEAUREGARD, DODO (Dodo), Mr.Anderson, and User.

\begin{figure}[!th]
    \centering
    \includegraphics[angle=90,origin=c,width=\textwidth]{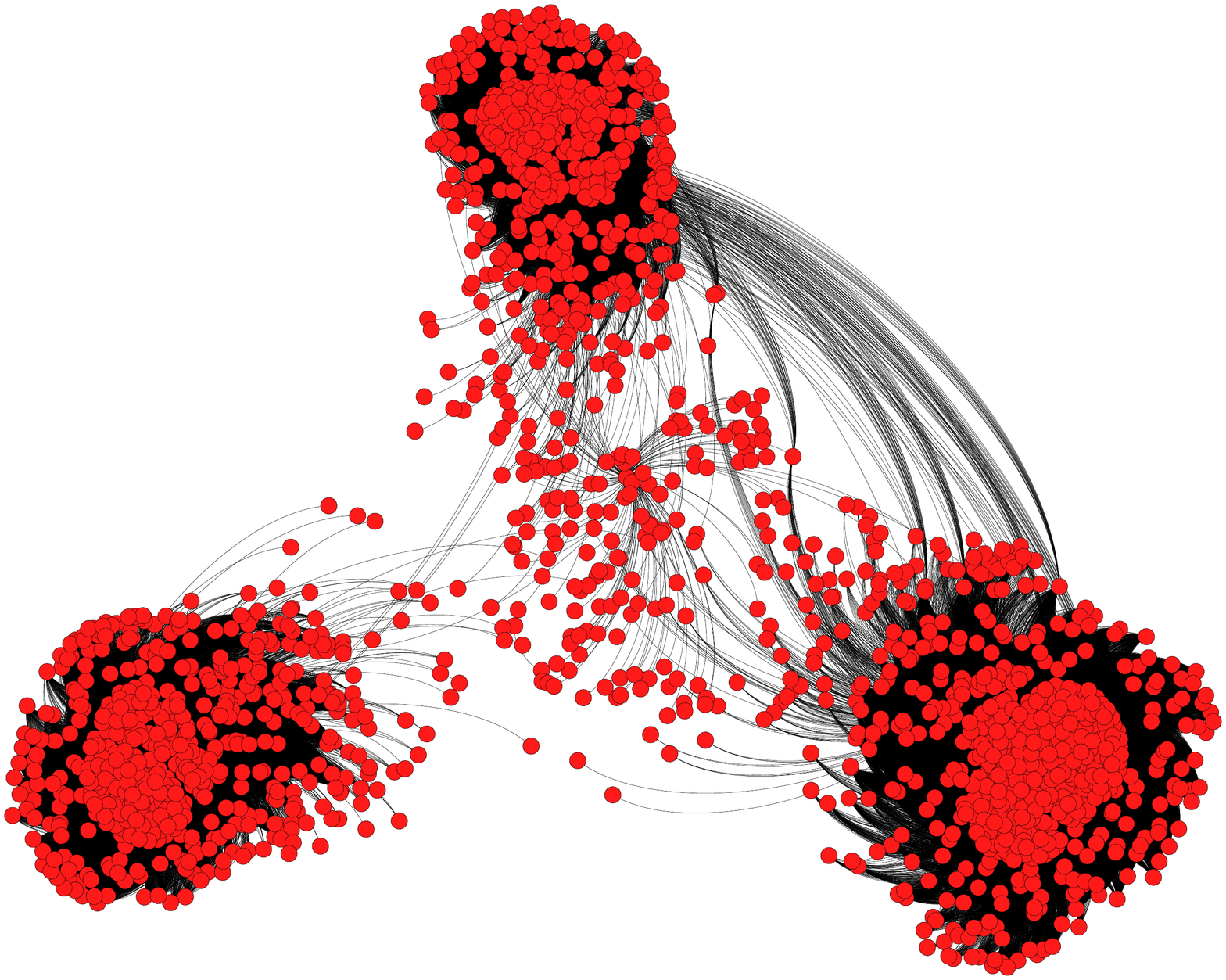}
    \caption{Epoch clustering of the collected samples.}
    \label{fig:epochs}
\end{figure}

To communicate the extracted data to the corresponding C2 server, each bot is using a custom protocol over HTTP. To encrypt the messages, they use AES with a random key of 128 bits. The key is encrypted using an RSA key which is 768 bits long and differentiates in the Epochs. From the analysed samples, we have collected four different RSA public keys. Using the RSA keys and the IP addresses of the C2 servers, one may cluster the samples in the three different epochs (see Figure \ref{fig:epochs}) and observe the interdependencies among the clusters, e.g. reused IP addresses of C2 servers.

The used ports of the C2 servers are 20, 80, 443, 990, 4143, 7080, 8080, 8081,8090 and 8443, with some of them using more than one port during the campaign. 

As discussed, upon host infection, Emotet will communicate with the C2 to fetch other malware in the form of an executable file. It should be noted that the executable, which is downloaded, depends on the country from which the request was made as the botnet serves different files to apply different ``agenda'' to each conducted campaign.

\section{Things to look for in future campaigns}
In future campaigns, blue teams have to be more cautious as several changes further extend the impact of the attack. To this end, we have highlighted the following aspects: 
\begin{itemize}
    \item Use of macroless malicious office documents, to exploit, e.g. using DDE.
    \item Different LOLBAS to launch the binaries of Emotet.
    \item Better obfuscation and increased evasion methods.
    \item More attacks to ViperMonkey and the like to impede the analysis of the documents.
    \item Use of DGAs to connect with the C2 servers.
    \item More targeted e-mails using, e.g. the native language of the potential victim.
\end{itemize}

\section{Conclusions and recommendations}
The infection rate of Emotet showcases several critical issues. Firstly, it is evident that the human factor and the lack of proper user training is the primary cause of this broad infection. Users must be trained not to open suspicious and out of context e-mails, even if originating from seemingly known entities. Moreover, they should be trained to check the origin of the sender of an e-mail and to timely report security breaches. Secondly, e-mail filters that are widely used are not as efficient as initially considered, and adversaries may easily bypass them using, e.g. an encrypted attachment. Moreover, the execution of macros must be disabled from Microsoft Office installations, and end-point security mechanisms must monitor and block the triggering of LOLBAS through Microsoft Office documents. Finally, the broader use of DMARC, DKIM and SPF may significantly reduce e-mail address spoofing and thus counter the potential infection and impact of such campaigns. 

\section*{Acknowledgements}
This work was supported by the European Commission under the Horizon 2020 Programme (H2020), as part of the projects \textit{YAKSHA} (Grant Agreement no. 780498), \textit{LOCARD} (Grant Agreement no. 832735), and \textit{CyberSec4Europe} (Grant Agreement no. 830929).

The content of this article does not reflect the official opinion of the European Union. Responsibility for the information and views expressed therein lies entirely with the authors.


\bibliographystyle{unsrt}
\bibliography{references}

\end{document}